\documentclass[journal=jacsat,manuscript=article]{achemso}
\usepackage[version=3]{mhchem} 
\usepackage[utf8]{inputenc}
\usepackage[T1]{fontenc}
\usepackage{graphicx}
\usepackage{siunitx}
\UseRawInputEncoding 

\title{Direct observation of ultrafast lattice distortions during exciton-polaron formation in lead-halide perovskite nanocrystals}

\author{H\'el\`ene Seiler}
\affiliation{Fritz Haber Institute of the Max Planck Society, Faradayweg 4-6, 14195 Berlin, Germany}
\email{seiler@fhi-berlin.mpg.de}

\author{Daniela Zahn}
\affiliation{Fritz Haber Institute of the Max Planck Society, Faradayweg 4-6, 14195 Berlin, Germany}

\author{Victoria C. A. Taylor}
\affiliation{Fritz Haber Institute of the Max Planck Society, Faradayweg 4-6, 14195 Berlin, Germany}

\author{Maryna I. Bodnarchnuk}
\affiliation{Laboratory for Thin Films and Photovoltaics, Swiss Federal Laboratories for Materials Science and Technology, \"Uberlandstrasse 129, CH-8600 D\"ubendorf, Switzerland}

\author{Yoav W. Windsor}
\affiliation{Fritz Haber Institute of the Max Planck Society, Faradayweg 4-6, 14195 Berlin, Germany}
\alsoaffiliation{Institut f\"ur Optik und Atomare Physik, Technische Universit\"at Berlin,
Strasse des 17. Juni 135, 10623 Berlin, Germany}

\author{Maxsym V. Kovalenko}
\affiliation{Institute of Inorganic Chemistry, Department of Chemistry and Applied Biosciences, ETH Z\"urich, CH-8093
Z\"urich, Switzerland.
}
\alsoaffiliation{Laboratory for Thin Films and Photovoltaics, Swiss Federal Laboratories for Materials Science and Technology, \"Uberlandstrasse 129, CH-8600 D\"ubendorf, Switzerland}
\author{Ralph Ernstorfer}
\affiliation{Fritz Haber Institute of the Max Planck Society, Faradayweg 4-6, 14195 Berlin, Germany}
\alsoaffiliation{Institut f\"ur Optik und Atomare Physik, Technische Universit\"at Berlin,
Strasse des 17. Juni 135, 10623 Berlin, Germany}
\email{ernstorfer@fhi-berlin.mpg.de}

\begin{document}

\maketitle

\begin{abstract}
The microscopic origin of slow carrier cooling in lead-halide perovskites remains debated, and has direct implications for applications. Slow carrier cooling has been attributed to either polaron formation or a \textit{hot}-phonon bottleneck effect at high excited carrier densities (> 10$^{18}$ cm$^{-3}$). These effects cannot be unambiguously disentangled from optical experiments alone. However, they can be distinguished by direct observations of ultrafast lattice dynamics, as these effects are expected to create qualitatively distinct fingerprints. To this end, we employ femtosecond electron diffraction and directly measure the sub-picosecond lattice dynamics of weakly confined CsPbBr$_3$ nanocrystals following above-gap photo-excitation. The data reveal a light-induced structural distortion appearing on a time scale varying between 380 fs to 1200 fs depending on the excitation fluence. We attribute these dynamics to the effect of exciton-polarons on the lattice, and the slower dynamics at high fluences to slower hot carrier cooling, which slows down the establishment of the exciton-polaron population. Further analysis and simulations show that the distortion is consistent with motions of the [PbBr$_3$]$^{-}$ octahedral ionic cage, and closest agreement with the data is obtained for Pb-Br bond lengthening. Our work demonstrates how direct studies of lattice dynamics on the sub-picosecond timescale can discriminate between competing scenarios, thereby shedding light on the origin of slow carrier cooling in lead-halide perovskites. \end{abstract}

Lead halide perovskites (LHPs) have attracted significant attention for their remarkable opto-electronic properties, in particular their unusual photovoltaic performance \cite{Zhang2016, Protesescu2015, Sutherland2016, CorreaBaena2017}. There is ongoing debate over the origin of long carrier lifetimes observed in LHPs, which is of direct relevance to applications. One explanation is screening by large polaron formation, which may protect carriers from scattering by phonons and defects \cite{Zhu2016, Niesner2016,Miyata2017,Buizza2021}. At high excitation densities (> 10$^{18}$ cm$^{-3}$), a \textit{hot}-phonon bottleneck effect has also been considered to explain the observed slower carrier cooling rates. In such a scenario, a strongly nonthermal population of LO phonons generated by electron-phonon coupling remains out-of-equilibrium with other phonons for several picoseconds \cite{Yang2015, Price2015, Yang2017,Papagiorgis2017, Mondal2018, Fu2017, Butkus2017, Chan2021, Nie2020, Shi2020, Verkamp2021, Chen2019, Monahan2017}. These two scenarios are expected to give rise to qualitatively different lattice dynamics, and can therefore be distinguished by such observations. Hence having direct experimental access to the lattice dynamics of LHPs can enable elucidating the microscopic origin of the slow carrier dynamics in LHPs.\par
Time-resolved diffraction techniques are ideally-suited for this task. They offer the most direct measurement of nonthermal phonon populations in photo-excited materials \cite{Stern2018,Zahn2020, Seiler2021}. Recently, they have also emerged as powerful methods to probe polaronic effects \cite{Guzelturk2021, Seiler2021b, RendeCotret2022}. Several time-resolved diffraction studies have already reported light-induced lattice dynamics of the soft lattice in LHPs \cite{Wu2017,Kirschner2019,Guzelturk2021, Cannelli2021}. Femtosecond electron diffraction (FED) was successfully employed to monitor the formation of a rotationally disordered halide octahedral structure over several picoseconds \cite{Wu2017}. More recently, time-resolved X-ray diffuse scattering revealed transient strain fields building over tens of picoseconds after polaron formation \cite{Guzelturk2021}. Using time-resolved X-ray absorption spectroscopy, Cannelli and co-workers were able to identify the photo-induced polaronic distortion of the lattice tens of picoseconds after photo-excitation \cite{Cannelli2021}. These works clearly demonstrate the benefits of direct structural probes of the soft LHP lattice. However, while these studies have mainly focused on processes on several picosecond timescales, investigating the sub-picosecond lattice dynamics is extremely relevant as well, as competition between hot-carrier thermalization and polaron formation is expected to occur on these timescales.\par

Here we employ femtosecond electron diffraction (FED) to probe the sub-picosecond lattice dynamics in weakly confined CsPbBr$_3$ nanocrystals (NCs) after photo-excitation above the electronic band gap. Both hot electron cooling via the lattice and polaron formation can be expected to occur under our excitation conditions \cite{Evans2018, Bretschneider2018}. The data directly reveal the emergence of a light-induced structural distortion, which builds up with a time constant ranging from 380 to 1200 femtoseconds depending on the excitation density (0.7 to 5.6 $\times$ 10$^{19}$ cm$^{-3}$). This observation is consistent with the establishment of an exciton-polaron population in the NCs. Combining structure factor analysis and simulations of diffraction patterns for distorted structures, we find that our data are qualitatively consistent with specific motions of the [PbBr$_3$]$^{-}$ octahedral cage, in particular Pb-Br$_2$ bond lengthening (see Figure \ref{fig1}(a)). Furthermore, all the observables in our data are well-modeled by a similar sub-picosecond time constant. The fluence dependence of this sub-picosecond time constant can be explained by slower hot carrier cooling at high fluences, reported in several previous studies \cite{Hopper2018, Price2015, Verkamp2021,Diroll2019}. These results suggest hot electron cooling and the creation of an exciton-polaron population occur in a coupled fashion. Our work demonstrates the value of measuring the lattice dynamics directly to probe the interplay of the various competing effects at the origin of long carrier lifetimes in LHPs. 

\section{Results and Discussion}
\begin{figure*}[ht!]
    \centering
    \includegraphics[width=0.9 \linewidth]{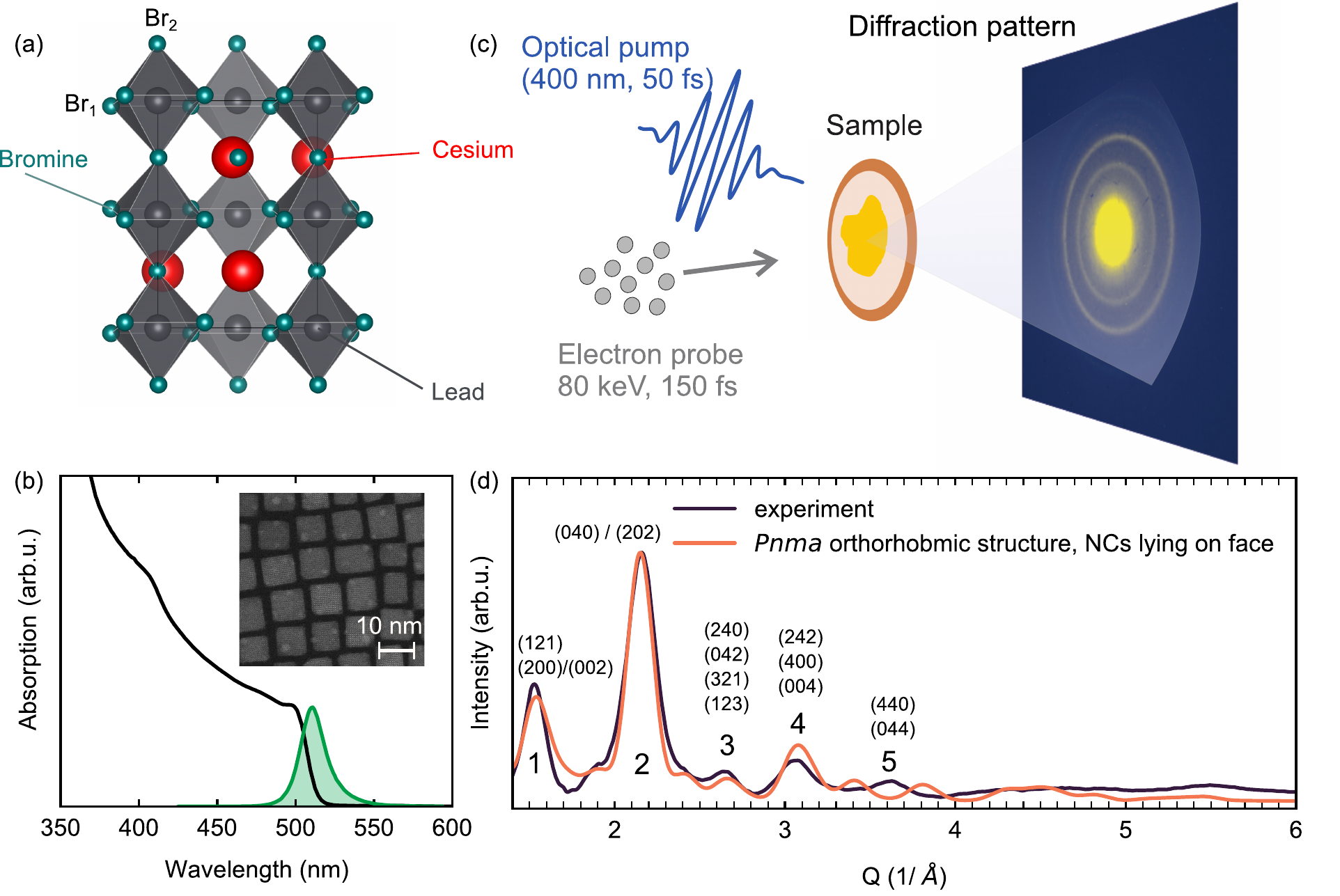}
    \caption{(a) Orthorhombic crystal structure of CsPbBr$_3$ from Ref. \citenum{Stoumpos2013}, with the two inequivalent Bromine atoms labeled. (b) Linear absorption (plain black line) and photoluminescence (filled green) spectra of the CsPbBr$_3$ NCs dispersed in toluene. Inset: TEM picture showing the NCs in real space. (c) Schematic illustration of the FED experiment, with an example diffraction pattern of the NCs as collected by our detector. (d) Diffraction profile of the CsPbBr$_3$ NCs (dark line), obtained by azimuthally averaging the pattern shown in (c). An empirical function was employed to remove background contributions. The orange curve represents the simulated pattern using the structure from Ref. \citenum{Stoumpos2013} and assuming the NCs lie flat on their faces.}
    \label{fig1}
\end{figure*}

CsPbBr$_3$ nanocrystals were synthesized following previously established procedures \cite{Protesescu2015, Bodnarchuk2018}. The linear absorption spectrum of the NCs dispersed in toluene is shown as the black curve in Figure \ref{fig1}(b), featuring a bandgap of 2.5 eV (496 nm). The inset of this panel shows a representative TEM image of the nanocrystals. The size of the nanocrystals is $\simeq$ 10 nm, indicating weak quantum confinement effects, since the exciton Bohr radius is $\simeq$ 7 nm for CsPbBr$_3$ \cite{Protesescu2015}. The linear photoluminescence spectrum, shown as the solid green line in Figure \ref{fig1}(b), is redshifted by a Stokes shift of about 30 meV. \\

Following basic optical characterisation of the samples, the NCs were drop-casted on a 10 nm thick Quantifoil TEM membrane (Plano GmbH) for the FED measurements. The NCs film thickness is estimated to be around 60 nm based on transmission measurements performed in an optical microscope with a narrow bandpass filter at 400 nm and previously determined values of intrinsic absorption coefficients in CsPbBr$_3$ nanocrystals \cite{Maes2018}. An example of an equilibrium transmission electron diffraction pattern of the perovskite NCs is presented in Figure~\ref{fig1}(c). Due to averaging over a wide range of orientations of the NCs probed by the electron beam, the diffraction pattern exhibits Debye-Scherrer rings typical of polycrystalline samples. For further analysis, the diffraction pattern is azimuthally averaged and the inelastic background arising from the substrate is removed (see supplementary Figure~1). An azimuthally averaged and background-subtracted diffraction profile is shown in Figure \ref{fig1}(d).\par

The thermal equilibrium structure of perovskite NCs is characterized by a complex structural landscape, featuring local polar fluctuations among different noncubic structures \cite{Yaffe2017}, significant local distortions of the PbX$_6$ octahedra \cite{Worhatch2008}, structural defects and twin boundaries \cite{Bertolotti2017}. We find that the experimental pattern in Figure~ \ref{fig1}(d) is best reproduced by simulating the pattern for the $Pnma$ orthorhombic structure, assuming that the NCs lie on one of their faces \cite{Stoumpos2013} (see Supplementary Figure 2). The simulated pattern is shown as the orange curve in Figure~\ref{fig1}(d). Within the limit of the coherence length of our electron beam, the positions of the Bragg reflections in our measured diffraction pattern are consistent with the simulated pattern as well as previous experimental studies \cite{Kirschner2019, Cannelli2022}. The Miller indices corresponding to the peaks are labeled in Figure~\ref{fig1}(d). In the remainder of this work we will refer to the peaks as 1-5 for convenience. \par
\subsection{FED results}
FED was previously applied successfully to other types of NCs \cite{Wang2016, Mancini2016, Vasileiadis2018, Vasileiadis2019, Krawczyk2021}. A schematic illustration of the experiment is shown in Figure~\ref{fig1}(c): a femtosecond laser pulse is used to impulsively excite the electrons in the material. After a controllable time delay $t$, an electron pulse diffracts off the lattice. The resulting diffraction pattern encodes the non-equilibrium state of the lattice at $t$. By varying the time delay between the pump and the probe, the ultrafast lattice dynamics following photo-excitation can be monitored. Further details about the FED instrument are available elsewhere~\cite{Waldecker2015}. Here, the CsPbBr$_3$ NCs are photo-excited with a 50-fs light pulse with central photon energy h$\nu$ = 3.1 eV (400 nm), roughly 0.6 eV above band edge. All measurements are performed at room temperature. The incident fluence on the sample is varied in the range from 0.09 to 0.70 mJ/cm$^2$, and the resulting initial density of photo-excited carriers induced by the pump pulse is estimated to be in the range from n$_{\rm e}$ = 0.7 - 5.6$\times$ 10$^{19}$ cm$^{-3}$ (see Supporting Information). At these carrier densities, we estimate that each NCs hosts multiple excited charge carriers (see Supporting Information). After photo-excitation of the CsPbBr$_3$ NCs, excited electrons and holes are generated. These hot charge carriers subsequently couple to the lattice via electron-phonon coupling, and we follow the ensuing lattice dynamics by investigating changes in the diffraction patterns as a function of pump-probe delay. \par

\begin{figure*}[ht!]
    \centering
    \includegraphics[width=1 \linewidth]{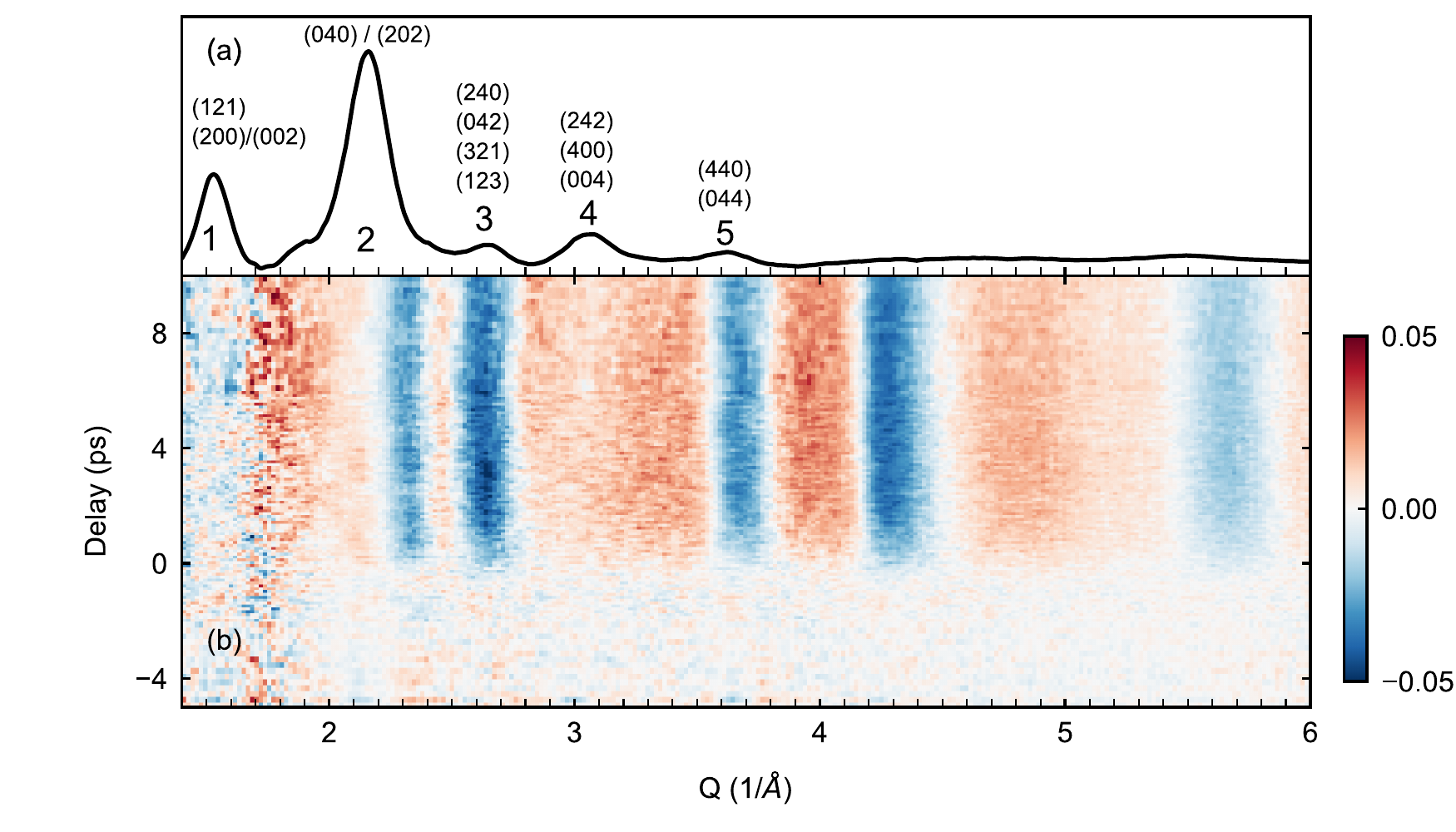}
    \caption{(a) Same as in Figure \ref{fig1}(d), reproduced for convenience. (b) Relative intensity difference map, shown here for an excitation density of 2.8$\times$ 10$^{19}$ cm$^{-3}$.}
    \label{fig2}
\end{figure*}
Figure~\ref{fig2} presents an overview of the photo-induced lattice dynamics, in the form of relative intensity difference maps. These difference maps are obtained as [I(t) - I(t<t$_0$)]/I(t<t$_0$), where I(t) is the diffraction profile at time delay t and t$_0$ is time zero. As shown in Supplementary Figure 3, the observed lattice dynamics remain qualitatively the same for all measured excitation densities. We verified that no time-resolved signal could be detected from the Quantifoil substrate (Supplementary Figure 4) under the same experimental conditions. In addition, the observed dynamics are reproducible over multiple scans acquired at different laboratory times (Supplementary Figure 5). The data in Figure 2 reflect complex lattice dynamics in addition to simple lattice heating. The latter was estimated to be only about 2 K for an excitation density of 2.8$\times$10$^{19}$ cm$^{-3}$ (see Supporting Information). Thermal heating leads to an intensity decrease of all Bragg peaks as per the Debye-Waller effect, see for instance Ref.~\citenum{Zahn2021}. Such a response is clearly not observed here for peaks 1, 2 and 4. Furthermore, peaks 2 and 5 shift to a lower scattering vector after photo-excitation, while other peaks do not follow this behaviour. Hence the data is also inconsistent with simple thermal expansion, where all peaks would go to lower $Q$ vectors. This simple overview of the data therefore suggests that the photo-induced lattice dynamics reflect some more complex light-induced structural distortion arising from electron-phonon interactions.\par

\begin{figure*}[ht!]
    \centering
    \includegraphics[width=\linewidth]{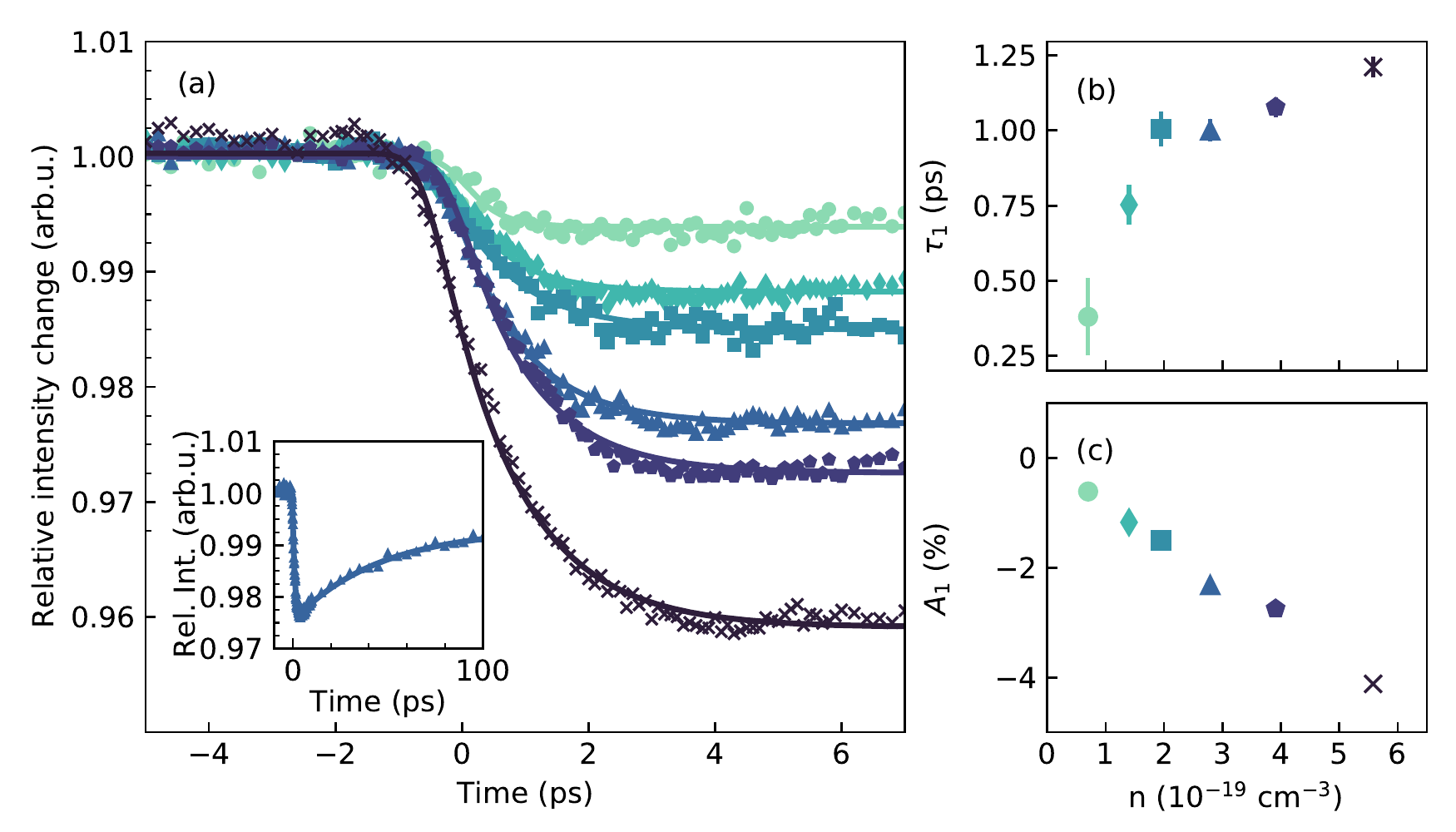}
    \caption{(a) Time-resolved relative diffraction intensities of the CsPbBr$_3$ NCs for various excitation densities, obtained by averaging the raw diffraction signals over some regions of interest (see Supplementary Figure 6 for more details on how they were obtained). The color code is matched to that of panels (b) and (c), which display the values of the corresponding excitation densities on their x-axis. Inset: example of a time-resolved trace over the 100 ps time range. Following the drop in intensity, the subsequent recovery indicates the onset of lattice cooling to the substrate.  (b) Time constant $\tau_1$ extracted from a bi-exponential fit to the data as a function of excitation density. The errors correspond to 68\% confidence intervals of the fits. (c) Amplitude $A_1$ extracted from the same fit as a function of excitation density.}
    \label{fig3}
\end{figure*}

Figure~\ref{fig3} shows the time-resolved relative diffraction intensities of the CsPbBr$_3$ NCs for various excitation densities, obtained by averaging the raw diffraction signals over the regions of interest (ROIs) shown in Supplementary Figure 6. All the ROIs exhibit the same dynamic response. An extended time range is presented in the inset of panel (a). The transient diffraction intensity can be fitted to a bi-exponential function convolved with a Gaussian (FHWM of 300 fs) to account for the finite temporal resolution of the experiment, see solid curves in Figure~\ref{fig3}(a). The fit results reveal that the lattice dynamics are well-captured by two time constants: a sub-picosecond time constant $\tau_1$ associated with the initial decrease in peak intensity, and a slow time constant $\tau_2$ of around 20 ps. We assign the slow time constant to heat transfer from the NCs to the Quantifoil substrate, and do not analyze it further. The fast time constant $\tau_1$ is intrinsic to the CsPbBr$_3$ NCs and reveals the reponse of the lattice to the excitation. Figures~\ref{fig3}(b) and (c) show the evolution of $\tau_1$ and the associated fit amplitude $A_1$ as a function of excitation density. We observe that $\tau_1$ rises with increasing excitation density, from 0.38 $\pm$ 0.13 ps at 0.7$\times$10$^{19}$ cm$^{-3}$ to 1.17 $\pm$ 0.03 ps at 5.6$\times$10$^{19}$ cm$^{-3}$. Meanwhile, the fit amplitude of the decay, $A_1$, increases from about 0.5$\%$ to around 4$\%$. This indicates, as expected, that the effect becomes more pronounced at high excitation densities.\par

To complement the analysis shown in Figure~\ref{fig3}, we determine the fluence dependence of the peak position variation of peak 2 (Supplementary Figure 7). Here also, we fit the peak position dynamics to a bi-exponential function convolved with a Gaussian. We observe similar values and trend for the fast time constant, $\tau_1^{\textrm{p2}}$, compared to those in Figure~\ref{fig3}(b). In contrast, no clear trend is seen for the amplitude of the peak shift, $A_1^{\textrm{p2}}$,  as a function of fluence. Together, the results of Figure~\ref{fig3} and Supplementary Figure 7 suggest all observables in the data (relative intensities and peak positions) follow the same sub-picosecond dynamics and do not reflect independent processes. In the Supplementary Information, we further show that such fast peak position changes do not violate speed of sound propagation in the specific case of NCs, owing to their high surface to volume ratio. \par

\subsection{Analysis of the structural distortion}
\begin{figure*}[ht!]
    \centering
    \includegraphics[width=1 \linewidth]{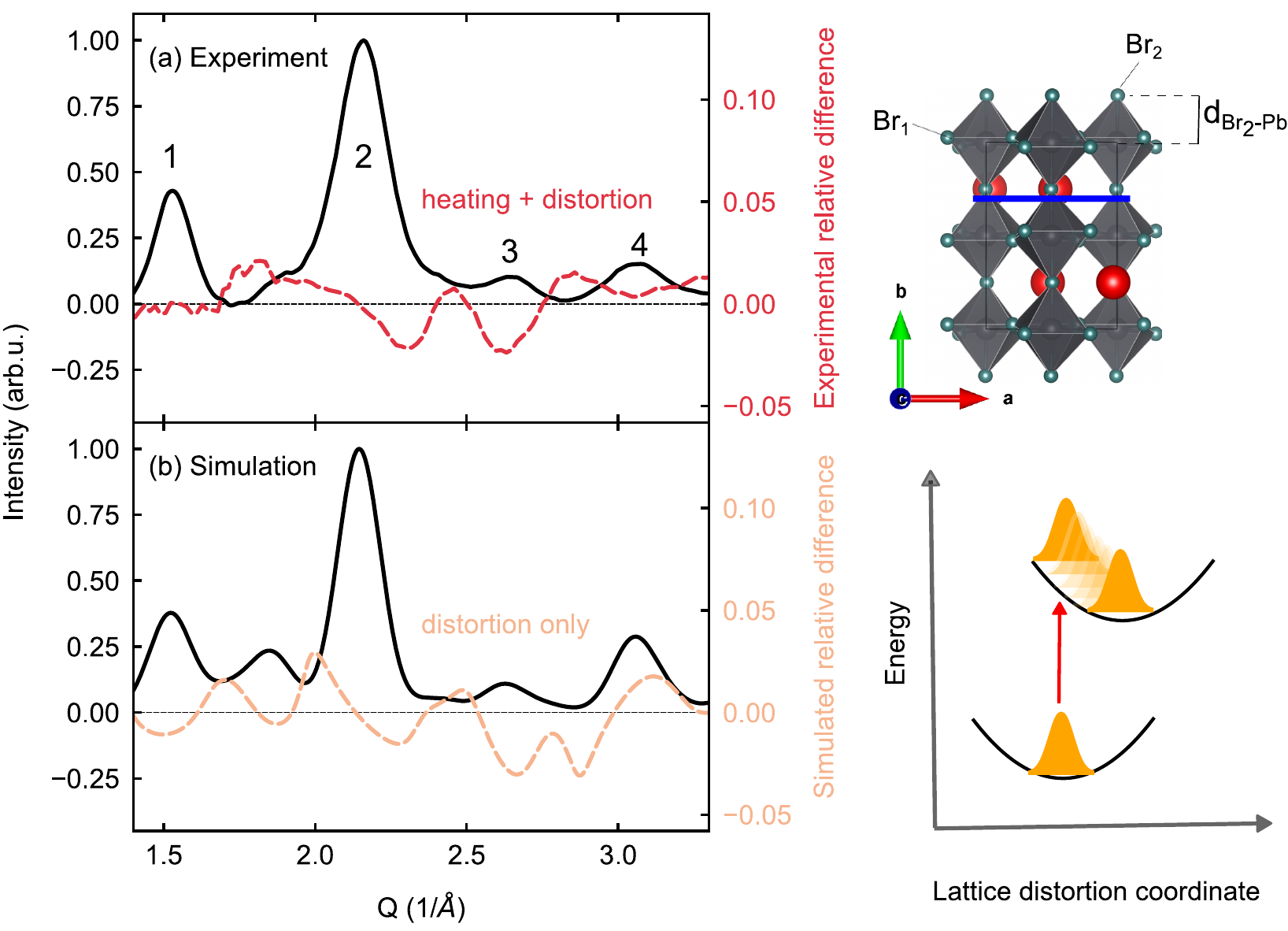}
    \caption{(a) Experimental diffraction profile of the CsPbBr$_3$ NCs (black), and relative difference profile from the experiment (dashed red). (b) Simulated diffraction profile of the CsPbBr$_3$ NCs (black), and simulated relative difference profile (dashed orange). More details about the distortion simulations are found in the text and Supplementary Information. (c) The (040) Miller plane is indicated (blue). The distortion simulated in panel (b) consists of a lengthening of Pb-Br$_2$ bond by 0.09 \%, estimated from the relative shift of peak 2. (d) Schematic illustration of the exciton-polaron formation process. The collective lattice dynamics following photoexcitation (red arrow) result in excited-state dynamics (orange wavepackets) on the excited potential energy surface that evolve from an initial state toward equilibrium.}
    \label{fig4}
\end{figure*}

We next evaluate possible real-space atomic motions at the origin of the structural distortion. We investigate commonly observed distortions in perovskites and whether they can give rise to the lattice dynamics in Figure~\ref{fig2} \cite{Woodward1997}. Specifically we consider tilting and distortions of the octahedra (e.g. changes in the Pb-Br bonds). Octahedral tilting, in particular, was reported to occur in response to ultrafast photo-excitation in other perovskites such as SrTiO$_3$ \cite{Porer2018, Porer2019}. For the analysis, we follow a similar approach as in Ref. \citenum{Seiler2021b}. We use the fact that atomic motions perpendicular to a lattice plane ($hkl$) modify the corresponding scattering intensity I$_{hkl}$, but in-plane motions do not. We start from peak 2, since it shows the clearest signature. Peak 2 is only sensitive to the (040) and (202) Miller planes, shown in Supplementary Figure 10. Having shown that the observed peak shift cannot be reproduced by intensity distribution changes between the (040) and (202) reflections, we list the possible atomic motions contributing to the signal. For the (040) plane, for instance, either a modification of the Pb-Br$_2$ bond or a tilting of the octahedra along the $c$ or $a$ axes of the crystal would change I$_{\textrm{040}}$, see Figure~\ref{fig4}(c). A similar reasoning can be applied to the (202) plane. The octahedra tilting angle or bond length changes can be estimated based on the shift of peak 2 at late delays (see Supplementary Information). Each possible distortion is individually simulated by modifying the unit cell according to these estimates, and diffraction patterns are generated for the modified structures. This procedure enables us to directly compare the simulated and experimental difference diffraction patterns for the different cases. Examples of a few distortions and simulated patterns are shown in Supplementary Figure 10. In Supplementary Figure 11, we also simulate a phase transition from the orthorhombic to the cubic phase, previously reported in a tr-XRD study on similar CsPbBr$_3$ NCs by the authors of Ref.~\citenum{Kirschner2019}.\par
The best agreement with the data is reached by a lengthening of the Pb-Br$_2$ bond, see Figure~\ref{fig4}(a-c). This distortion reproduces the peak shift of peak 2, the intensity reduction in peak 3 and the intensity rise of peak 4. The magnitude of the simulated relative difference is also in agreement with the experimental relative difference. Overall, the agreement remains qualitative due to heating effects being neglected (see Supporting information) and the sheer complexity of the LHP lattice structure. However, our work strongly suggests the involvement of Pb-Br cage motions in the build-up of the light-induced distortion, and in particular changes in the Pb-Br$_2$ bond. \par 

The presence of polarons in LHPs has been claimed by multiple complementary techniques, ranging from optical~\cite{Miyata2017, Bretschneider2018,Thouin2019, Lan2019-od,Allen1987, Seiler2019, Buizza2021} and photoemission spectroscopies~\cite{Puppin2020} to structural probes~\cite{Guzelturk2021, Cannelli2021}. Both the timescales and nature of the lattice dynamics observed here are consistent with the polaron formation picture \cite{Miyata2017, Evans2018, Seiler2019}. Furthermore, several studies have also suggested the involvement of [PbBr$_3$]$^{-}$ cage motions in polaron formation \cite{Miyata2017, Bretschneider2018, Park2018, Ambrosio2018,Schlipf2018-ke}, and atomic motion along the Pb-Br$_2$ direction \cite{Cannelli2021}. Figure \ref{fig4}(d) summarizes our interpretation of the data, in which lattice reorganization follows photoexcitation (red arrow), i.e. the lattice evolves from an initial state toward a new equilibrium. We note that for the investigated excitation densities here, each NC hosts several exciton-polarons whose radii may overlap~\cite{Frost2017}.  \par

\subsection{Interplay between hot carrier cooling and the creation of an exciton-polaron population}
In addition to the light-induced structural distortion, there are lattice heating contributions to the data arising from carrier cooling. We estimate that the Debye-Waller effect generates between 0.2 and 1$\%$ peak intensity losses depending on the scattering vector and excitation density (see Supporting Information). Thus, while heating may not dominate the lattice dynamics, it also cannot be neglected. Within our instrument response function of 300 fs, we do not observe hot carrier cooling and the emergence of an exciton-polaron population to occur in a two-step fashion. The peak shift dynamics - which can be assumed to reflect primarily the polaronic signatures - exhibit very similar time constants compared to the integrated ROIs, where lattice heating as a result of carrier cooling should clearly play a role. Therefore, our data suggest that hot carrier thermalization and exciton-polaron population build-up occur in a coupled fashion.\par

The increase of $\tau_1$ with increasing excitation density seen in Figure \ref{fig3}(b) and in Supplementary Figure 7(b) shows that the structural distortion exhibits longer time constants at higher fluences. We note that the intensity variations in our experiments reflect the population dynamics of exciton-polarons, which depend on both exciton-polaron formation and hot carrier cooling times \cite{Bretschneider2018}. Multiple studies have reported a slowing down of carrier cooling on the sub-picosecond timescale at high fluences \cite{Hopper2018, Price2015, Verkamp2021,Diroll2019}. Such trend could arise from carrier screening effects at high excitation densities, which are known to occur in polar semiconductors and would reduce the rate of phonon emission \cite{Yoffa1981}. Alternatively, from a simple two-temperature model, one would also expect an increase of the lattice heating time with increasing initial change in electronic temperature \cite{Allen1987, Anisimov1967}. Finally, the same trend would also be observed in the case of nonthermal phonon populations, which are likely present in our sample on the sub-picosecond timescale given the strong dependence of the time constant with fluence. Regardless of the origin of this dependence, the slower creation of the distortion at high fluences in our data is fully consistent with the slower hot carrier cooling rates observed by others. Further measurements pumping the NCs at the bandedge, where cooling effects are minimized, may isolate the exciton-polaron formation time in the future.\par

Even at the highest fluences, our measurements do not display signatures of lattice heating over a timescale of several picoseconds. At high excitation densities (> 10$^{18}$ cm$^{-3}$), several transient-absorption (TA) studies have reported slow components in the spectral dynamics, with time constants ranging from a few picoseconds \cite{Butkus2017, Mondal2018, Papagiorgis2017, Price2015, Nie2020} to tens or even hundreds of picoseconds \cite{Yang2015, Fu2017, Yang2017}. The interpretation of these slow components is controversial and lacks a commonly accepted picture \cite{Fu2017, Li2017, Butkus2017, Chan2021}, with some studies assigning the slow dynamics to the \textit{hot}-phonon bottleneck effect \cite{Yang2015}, other studies assigning them to Auger relaxation processes \cite{Li2017}. For our inorganic NCs, the time-resolved diffraction data do not exhibit signatures of a long-lived \textit{hot}-phonon bottleneck. 

Our study has revealed the sub-picosecond lattice dynamics of photo-excited CsPbBr$_3$ NCs. The main observation was the direct observation of a structural distortion within hundreds femtoseconds, which we assigned to the lattice signature of an emerging population of exciton-polarons. The distortion is consistent with atomic motions of the [PbBr$_3$]$^{-}$ cage. We further observed that the exciton-polaron population takes more time to build-up at high fluences, which we attributed to slower hot carrier cooling. No \textit{hot}-phonon bottleneck effect lasting several picoseconds was observed for the investigated excitation densities, which nearly reached the damage threshold of the NCs. Our study sheds light on the microscopic origins of slow carrier cooling in LHPs by helping discriminate between the competing scenarios proposed in the literature.

\newpage

\begin{acknowledgement}
This work was funded by the Max Planck-EPFL-Center for Molecular Nanoscience and Technology, the Max Planck Society, the European Research Council (ERC) under the European Union's Horizon 2020 research and  innovation program (Grant Agreement Number ERC-2015-CoG-682843), and partially by the Deutsche Forschungsgemeinschaft (DFG) - Projektnummer 182087777 - SFB 951. H.S.~acknowledges support by the Swiss National Science Foundation under Grant No.~P2SKP2\textunderscore184100.  V.T.~acknowledges financial support from the Alexander von Humboldt Foundation. M.I.B.~thanks for the support from the Swiss National Science Foundation (grant number 200021\textunderscore192308, project Q-Light). M. I. B.~ and M.V.K~acknowledge support by the Research and Innovation Foundation of Cyprus, under the "New Strategic Infrastructure Units-Young Scientists" Program, grant agreement number "INFRASTRUCTURES/1216/0004", Acronym "NANOSONICS". M. I. B.~ and M.V.K~ are grateful for the use of the Empa Electron Microscopy Center facilities. Y.W.W acknowledges funding from the DFG within the Emmy Noether program under Grant No. RE 3977/1, as well as the TRR 227 (Project A10). \end{acknowledgement}

\begin{suppinfo}
Simulated diffraction profiles and Bragg peaks assignment; Calculation of excitation density; Estimate of temperature rise; Estimate of the Debye-Waller effect; Estimate of average distance of atom to nanocrystal surface; Determination of octahedral tilt angle.

\end{suppinfo}

\newpage
\bibliographystyle{unsrt}
\bibliography{bibliography}

\end{document}


\flushbottom
\maketitle

\thispagestyle{empty}
\newpage

\section{Supplementary Text} 



\subsection{Simulated diffraction profiles and Bragg peaks assignment}

We simulate the diffraction peak positions and intensities of CsPbBr$_3$ with the \textit{CrystalMaker} software as well as Python scripts, using the crystallographic information from Ref. \citenum{Stoumpos2013}. The cell parameters are $a$ = 8.24 $\si{\angstrom}$, $b$ = 11.74 $\si{\angstrom}$ and $c$ = 8.20 $\si{\angstrom}$  (orthorhombic $pnma$). Supplementary Figure \ref{figs1} shows a comparison between the experimental pattern and various simulated radial diffraction profiles. Panel (a) shows the simulated pattern assuming a completely random orientations of the NCs, akin to a powder pattern. While most simulated peak positions seem to match experimental ones, peak intensities are not well reproduced. A better agreement of both peak positions and intensities is reached assuming that the NCs preferably lie on one of their faces, but with that face at random angle with respect to the substrate. To simulate this scenario, we compute the single crystal diffraction pattern assuming the electron beam is travelling along the $b$ (or $a$, $c$) axis of the crystal, and we rotate that pattern around its center from 0 to 2$\pi$. Then we find the linear combination of the three patterns (along the $a$, $b$ and $c$ directions) that best reproduces the experimental pattern. The results of this procedure are shown in panel (b). A much better agreement between the experimental and simulated peak positions and intensities is achieved compared to the powder pattern case. As a further point of reference, in panel (c) we also show a pattern similarly simulated but assuming a cubic structure based on the crystallographic information from Ref. \citenum{osti_1272881}. As can be seen from comparing panel (c) with panel (b), the measured diffraction pattern resembles more the orthorhombic phase, consistent with previous measurements at room temperature \cite{Stoumpos2013}. Finally, in panel (d) we report the powder pattern of PbBr$_2$, which we would expect if the sample was significantly degraded \cite{osti_1202200}. We observe a complete lack of agreement between the simulated pattern of PbBr$_2$ and the experimental pattern. \par
Based on the results in Supplementary Figure \ref{figs1}, we proceed to the assignment of peaks 1-8 to specific Miller indices assuming the orthorhombic structure. The Miller indices and diffraction intensities corresponding to peaks 1-5 are shown in tables 1-5 below. The intensities are scaled with respect to the max intensity I$_{\max}$, obtained for the ($040$) reflection. We note that in the reciprocal space region spanning 4 < Q < 6 [1/\si{\angstrom}], corresponding to peaks 6-8, there are more than 90 Bragg reflections with I/I$_{\rm max}$ ranging from 2-12 \%. Due to the large density of peaks, we do not proceed to an assignment of Miller indices in that region. 
\newpage

\subsection{Calculation of excitation density}
The aim is to determine the excitation densities in the perovskite NCs following laser excitation. We start by considering the photon energy of 3.1 eV (400 nm):
\begin{equation}
E_{\rm ph} = \frac{\rm hc}{\lambda} = \frac{2\times 10^{-25}\,\rm J\times m}{400 \times 10^{-9}\,\rm m} = 5 \times 10^{-19} \rm J  
\end{equation}
The incident number of photons per area is:
\begin{equation}
N_{\rm ph,in} = \frac{F_{\rm i}}{E_{\rm ph}} =  \frac{0.09 \times 10^{-3} \rm J/cm^{-2}}{5 \times 10^{-19} \rm J} \approx  1.8 \times 10^{14} \: \rm  cm^{-2},
\end{equation}
Where we take $F_i \simeq$ 0.09 mJ/cm$^{-2}$ as an example incident fluence on the sample. Next we calculate the absorbed number of photons per unit volume. The fraction of absorbed photons is $A$ = 1 - $T$ - $R$, where $T$ and $R$ are the transmission and reflection coefficients, respectively. We determine $A$ from transfer matrix calculations based on the complex index of refraction of CsPbBr$_3$ at 400 nm and the estimated film thickness of 60 nm \cite{Zhao2018, Yan2020}. Hence the number of absorbed photons per unit volume is: 
\begin{equation}
N_{\rm ph,abs} = \frac{A \times N_{\rm ph,in}}{d} = \frac{0.24 \times 1.75 \times 10^{14} \rm cm^{-2}}{60 \,\rm nm} \approx 7 \times 10^{18} \,\rm cm^{-3},
\end{equation}
Taking into account the volume of a NC, $V_{\rm NC} = 1 \times 10^{-18} \: \rm cm^3 $, then number of excitations per NC can then be roughly estimated by:
\begin{equation}
N = N_{\rm ph,abs}\times V_{\rm NC} = 7 \times 10^{18} \,\rm cm^{-3} \times 1 \times 10^{-18}\,\rm cm^3 \approx 7
\end{equation}
We see that at the high excitation densities employed in our study, we have a large number of excitations (N $\gg$ 1) per nanocrystal. Table \ref{table1} summarizes the incident fluences $F_i$ and the resulting excitation densities. 

\begin{center}
\begin{tabular}{ c c c c c c c}
F$_i$ [mJ/cm$^2$] & 0.09 & 0.17 & 0.24 & 0.35 & 0.49 & 0.70  \\ 
$N_{\rm ph,abs}$ [10$^{19}\times$ cm$^{-3}$] & 0.70 & 1.40 & 1.95 & 2.79 & 3.91 & 5.58   \\
\label{table1}
\end{tabular}
\end{center}
\newpage

\subsection{Estimate of temperature rise}
We estimate the temperature rise in the photo-excited perovskite NCs:
\begin{equation}
    \Delta T = \frac{E_{\mathrm{abs}}}{C}
\end{equation}
In this expression, $C \simeq$ 125 J/[mol$\times$ K] is the heat capacity of CsPbBr$_3$, from Ref. \citenum{Evarestov2020} and $E_{\mathrm{abs}}$ is the absorbed energy in Joule per mole. To find a value for $E_{\mathrm{abs}}$, we first calculate the excess energy per unit cell in a perovskite NC. The number of unit cells $N_c$ in a cubic NC of 10 nm size is $N_c = V_{\mathrm{NC}}/V_{\mathrm{uc}}$, where $V_{\mathrm{NC}} = 1000  \: \mathrm{nm}^3$ and $V_{\mathrm{uc}} = 0.85 \: \mathrm{nm}^3$ (using the unit cell volume of the conventional standard orthorhombic unit cell in Ref. \citenum{osti_1273967}). This yields $N_c \simeq 1180$. In the previous section, we estimated that each NC hosts multiple excitations, each involving 600 meV excess energy with respect to the bandgap at 2.5 eV. We neglect non-radiative recombination pathways, as the photoluminescence quantum yield of these NCs was determined to be 90 $\%$ \cite{Protesescu2015}. Hence, neglecting the electronic heat capacity, the absorbed energy per unit cell that ends up in the lattice is:
\begin{equation}
   E_{\mathrm{abs}} = \frac{N \times 600}{1180} = N \times 0.51 \frac{\mathrm{meV}}{\mathrm{unit\:cell}}
\end{equation}

Where N is the average number of excitations in a NC. Taking N = 23 (estimate corresponding to an excitation density of 2.79 $\times$ 10$^{19}$ cm$^{-3}$), we have 11.7 meV per unit cell excess energy. Each unit cell contains 12 Br atoms, 4 Cs atoms and 4 Pb atoms \cite{osti_1273967}. Hence, each of the four CsPbBr$_3$ has roughly 3 meV excess energy. We multiply this quantity by the Avogadro number to obtain the energy per mole:
\begin{equation*}
    E_{\mathrm{abs}} = 3 \times 10^{-3} \times 6.02 \times 10^{23} =  17.60 \times 10^{20} \: [\mathrm{eV/mol}] = 1.60\times10^{-19} \times 17.60 \times 10^{20} [\mathrm{J/mol}] \: \simeq \: 282 \: [\mathrm{J/mol}] 
\end{equation*}

Using the heat capacity of CsPbBr$_3$ of 125 J$\times$ mol$^{-1}\times$K$^{-1}$ from Ref. \citenum{Evarestov2020}, we can retrieve an estimated temperature rise of $\Delta T = 282/ 125 \simeq 2.25$ K \cite{Kittel2004}. Our estimated rise is in the same order of magnitude as estimated in a previous FED work on MaPbI$_3$ films \cite{Wu2017}. 
\newpage
\subsection{Estimate of the Debye-Waller effect}
We estimate the changes of the Debye-Waller (DW) factor that arise from a temperature increase $\Delta T$, as well as its impact on relative diffraction intensities. We do so using two independent methods. In both cases the aim is to give an order of magnitude for the DW effect as opposed to a quantitative value, which goes beyond the scope of this work. In the first method, we use experimentally determined values of the atomic mean-square displacement (MSD) parameters reported in Ref. \citenum{Lpez2020}. From this study, the equivalent isotropic displacement parameters at 293 K for the Cs, Pb and Br atoms are U$_\textrm{Cs}$ = 0.084 \AA$^2$, U$_\textrm{Pb}$ = 0.026 \AA$^2$, U$_{\textrm{Br}_1}$ =  0.086 \AA$^2$ and U$_{\textrm{Br}_2}$ = 0.071 \AA$^2$, respectively. The average displacement is then given by U$_{\textrm{av}}$ = 3/5$\times$(U$_\textrm{Cs}$ + U$_\textrm{Pb}$ + U$_{\textrm{Br}_1}$ + 2$\times$ U$_{\textrm{Br}_2}$) = 0.2 \AA$^2$, where the factor 3 accounts for the three independent spatial coordinates along which the atoms can vibrate. In the high-temperature limit, which is a reasonable assumption at room temperature, the MSD is proportional to the temperature. Hence we can estimate the MSD difference arising from a rise in temperature $\Delta T$ as:

\begin{equation}
\Delta \langle u \rangle ^2 = \langle u \rangle^2( \mathrm{T} = 295 \: K + \Delta \mathrm{T}) - \langle u \rangle^2 (\mathrm{T} = 295 \: K) = \frac{\Delta \mathrm{T}}{293 \: K}\times U_{\textrm{av}}.
\end{equation}

Where $\langle u \rangle ^2$ is the MSD. For $\Delta $T = 2.25 K, we get $\Delta \langle u \rangle ^2 \approx 1.6 \times 10^{-3}$ \AA$^2$. We now determine how much this MSD change impacts the relative Bragg peak intensities. For the (040) reflection (peak 2), we find that $I_{040}$(T = 297.25 K)/$I_{040}$(T = 295 K) = $e^{-\frac{1}{3}\Delta \langle u \rangle^2  |\vec{G}_{040}|^2} \approx $ 0.9976, with $|\vec{G}_{040}| \approx 2.14 \times 10^{10}\,{\rm m}^{-1}$. For the (044) reflection (peak 5), we find that $I_{044}$(T = 297.25 K)/$I_{044}$(T = 295 K) = $e^{-\frac{1}{3}\Delta \langle u \rangle^2  |\vec{G}_{044}|^2} \approx $ 0.9928, with $|\vec{G}_{044}| \approx 3.74 \times 10^{10}\,{\rm m}^{-1}$. This means that heating effects are expected to have a non-negligible but also not dominating contribution to the observed structural dynamics. 

For the second estimate of the DW factor, we consider a simple Debye model with a Debye temperature of 102 K, as obtained from literature for CsPbBr$_3$ \cite{Evarestov2020}. We calculate $\langle u^2 \rangle$ in thermal equilibrium at 300 K using the expression from Ref. \citenum{Peng:2004}:
\begin{equation} \added[]{
\langle u^2 \rangle = \frac{3\hbar}{2m} \int_0^{\omega_D} \coth \frac{\hbar\omega}{2k_{\rm B}T} \, \frac{g(\omega)}{\omega} d\omega\, ,}
\end{equation}

with $g(\omega)$ the density of phonon states from the Debye model, $\omega_D$ is the Debye frequency, $k_B$ is the Boltzmann constant and $m$ is the average atomic mass of CsPbBr$_3$. We obtain $\langle u^2 \rangle_{300\,\rm K} \approx 0.11$ \AA$^2$ or $\sqrt{\langle u^2 \rangle} \approx 0.33$ \AA.
The MSD change for $\Delta T \approx 2.25$ K retrieved is:
\begin{equation}
\Delta \langle u \rangle ^2 = \langle u \rangle^2( \mathrm{T} = 297.25 \mathrm{ K}) - \langle u \rangle^2 (\mathrm{T} = 295 \mathrm{ K}) \approx 9 \times 10^{-4} \textrm{\AA}^2.
\end{equation}

With this method, we find that $I_{040}$(T = 297.25 K)/$I_{040}$(T = 295 K) $\approx$  0.9986. For the (044) reflection (peak 5), we find that $I_{044}$(T = 297.25 K)/$I_{044}$(T = 295 K) $\approx$ 0.9958. This estimates yields smaller heating effects compared to the estimated based on the U-matrix formalism. Nevertheless, they confirm that heating effects are expected to play a non-negligible role, albeit not a dominant one, in the observed structural dynamics.

\newpage
\subsection{Estimate of average distance of atom to nanocrystal surface}
Assuming a cubic NC with an edge length of 10 nm and a unit cell length of 0.587 nm (the cubic crystal structure is considered here for the sake of simplicity), there are around 17 unit cells per edge. In total, the number of unit cells in the NC is given by 17$^3$ = 4913. We now build a NC layer by layer, starting from the center unit cell. For example, the layer surrounding the center unit cell has an edge length of $N$ = 3 unit cells, see Supplementary Figure 8. The number of unit cells for this layer, $N_l(3)$, can be easily shown to be 26. The counting is graphically illustrated in Supplementary Figure 8 as cubes of different colours. Based on this example, a similar counting procedure can be applied for the next layers, i.e. $N$ = 5, 7,... . One can easily show that the number of unit cells in a layer with an edge containing $N \geq $ 3 unit cells is given by $N_l (N) = 2\times N\times N \times  + 2\times N\times(N - 2) + 2\times(N-2)\times(N-2)$. We verify that the sum of the unit cells over all the layers in the NC gives the right number of unit cells:
\begin{equation}
     N_c = 1 + \sum_{N=3}^{17}N_l(N) = 4913
\end{equation}
With $N$ in the sum an odd number. By taking the ratio $N_l(N)/N_c$, one obtains the ratio of unit cells in the NCs belonging to a given layer. For instance, the ratio of unit cells at the surface is $N_l(17)/N_c = 0.31$.\par
Thanks to the ratios of unit cells in each layer, we determine the average distance of an atom in the NC to the surface to be 0.97 nm. Taking the speed of sound in CsPbBr$_3$ as $v_s = 1361$ m/s \cite{Elbaz2017}, the average time it would take for sound to propagate to the surface is given by $0.97\times10^{-9}/1361 \simeq 7 \times 10^{-13}$ s. The average time of roughly 700 fs is on the same order of magnitude as the timescale observed in the experiments. This calculation provides us with an independent estimate that it is indeed possible to observe unit cell changes in the NCs over such fast timescale. This is due to the high surface to volume ratio of nanoparticles.
\newpage
\subsection{Determination of octahedral tilt angle}
The aim is to estimate the tilt angle $\alpha$ as a function of $l$, both indicated in Supplementary Figure 9(a). Here we consider the case of a tilt along the $b$-axis as an example. We assume that the Br-Br-Br angles are 90$^{\circ}$. By definition we have $l/2 = \sqrt{2}$b$_r$, where b$_r$ = 4.15$\times10^{-10}$ m. We can write:
\begin{equation}
    \frac{l(\alpha)}{2\sqrt{2}\textrm{b}_r} = \textrm{cos} \beta
\end{equation}
Since $\beta = (\pi/2 - \alpha) -  \pi/4 = \pi/4 - \alpha$, we obtain:
\begin{equation}
\alpha(l) = \pi/4 - \textrm{acos} \Bigg [ \frac{l}{2\sqrt{2}\textrm{b}_r} \Bigg ]
\end{equation}
Here $l$ can be estimated from the shift of peak 2. A shift of 0.1 \%, for example, yields $l = l_{\textrm{ortho}}\times(1 + 0.001)$, where $l_{\textrm{ortho}}$ is the value of $l$ for the orthorhombic structure. In our case we find $\alpha \simeq 0.5^{\circ}$. Supplementary Figure 9(b) shows $\alpha$ as a function of $l/l_{\textrm{ortho}}$.

\newpage
\section{Supplementary Figures}
\begin{figure}[th!]
\renewcommand{\figurename}{Supplementary Figure}
\centering
\includegraphics[width=0.9\linewidth]{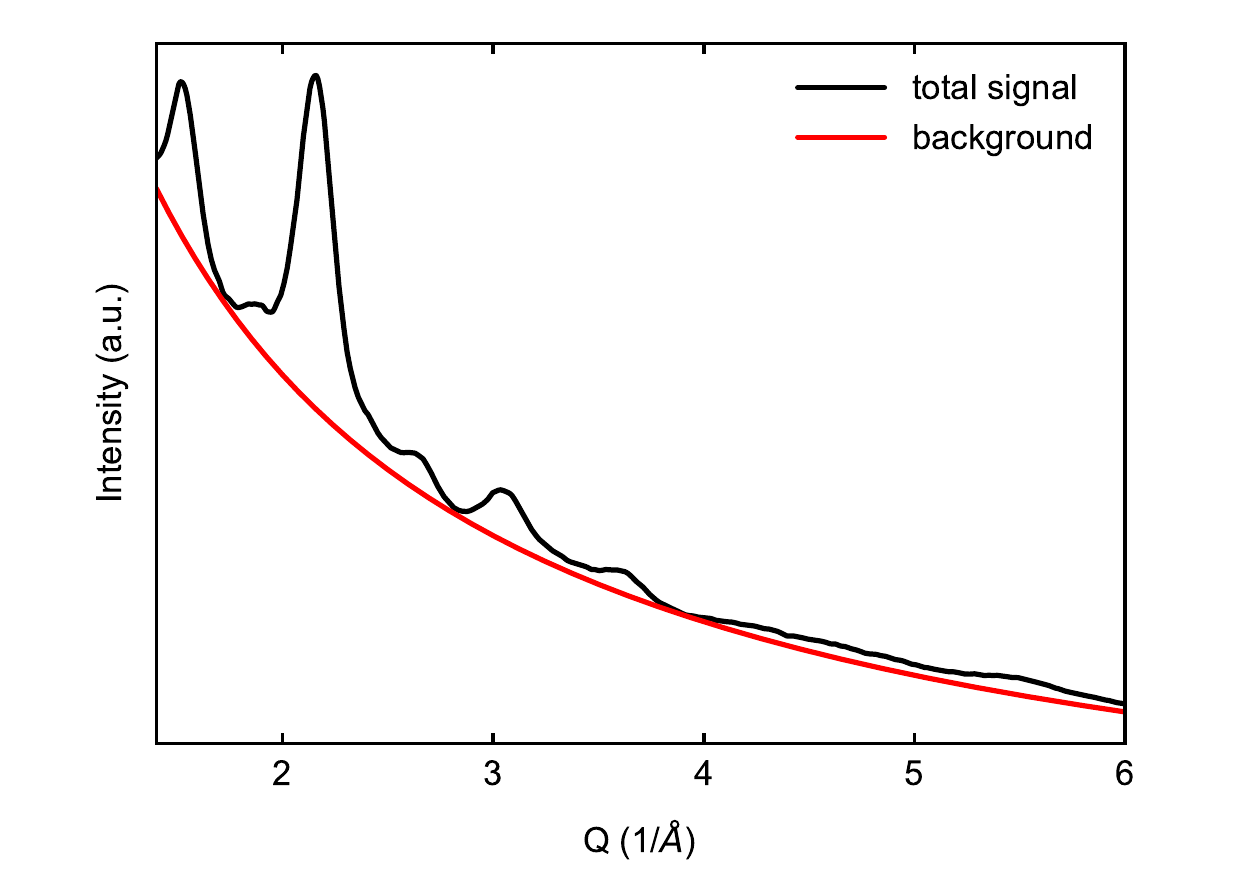}
\caption{Removal of background signals (red curve) arising from the substrate, diffuse scattering and contributions from the 0$^\textrm{th}$ order beam. Here, a Lorentzian tail and a constant offset were used as background function. We note that the qualitative features of Figure 2 in the main text remain the same whether the background-removed signals are employed or not.}
\label{figs0}
\end{figure}

\newpage
\begin{figure}[th!]
\renewcommand{\figurename}{Supplementary Figure}
\centering
\includegraphics[width=0.9\linewidth]{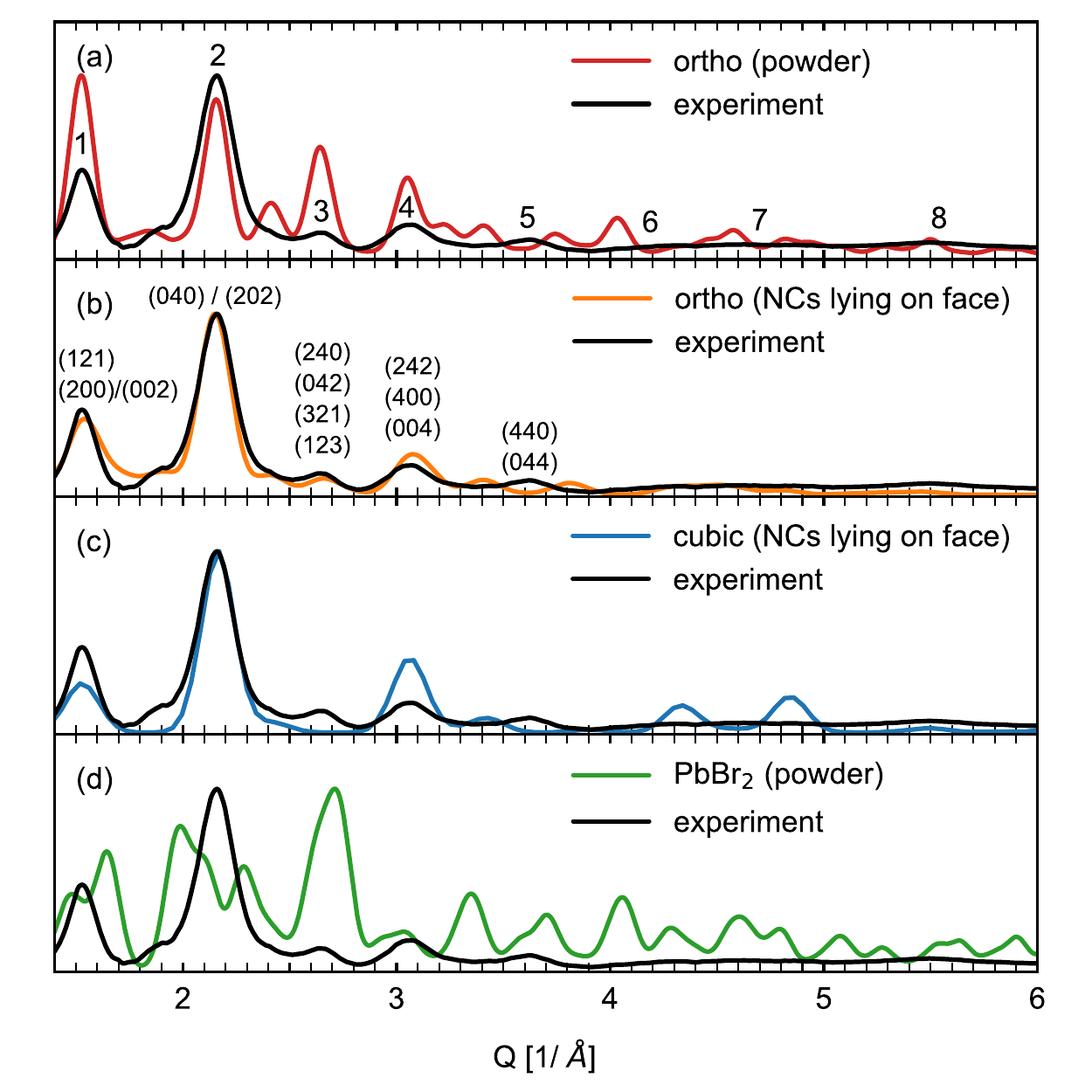}
\caption{Comparison of the experimental diffraction signal with various simulated patterns. (a) Simulated pattern assuming a powder-like distribution of NCs and an orthorhombic structure. (b) Simulated pattern for an orthorhombic structure, assuming the NCs lie on one of their faces on the substrate, but at random angles within the planes of the faces. More details are provided in the supplementary text. (c) Simulated pattern assuming a cubic structure, with the NCs lying on one of their faces on the substrate. (d) Simulated pattern of a PbBr$_2$ powder, expected for a significantly degraded sample.}
\label{figs1}
\end{figure}

\newpage

\begin{figure}[h!]
\renewcommand{\figurename}{Supplementary Figure}
\centering
\includegraphics[width=0.7\linewidth]{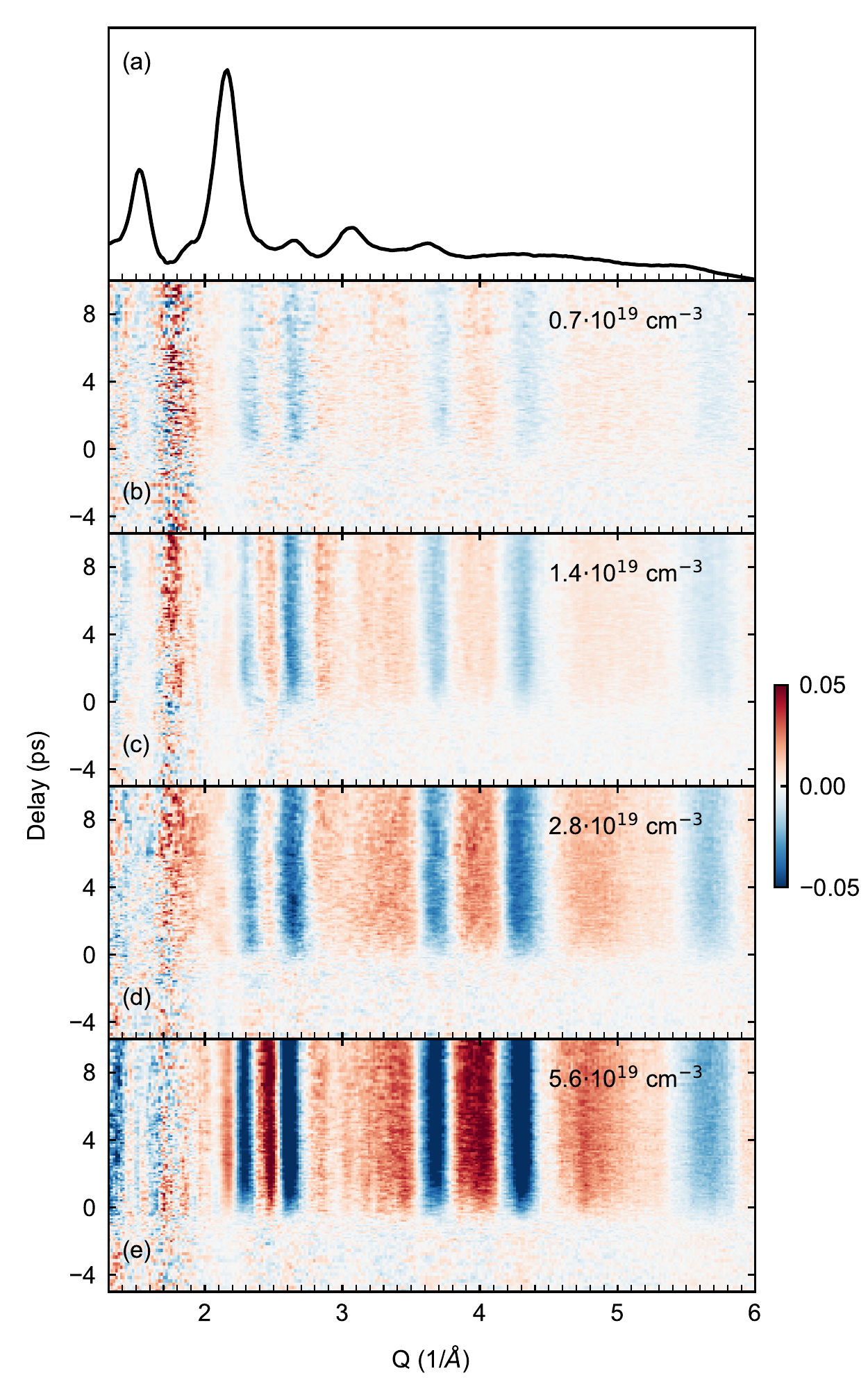}
\caption{(a) Background-substracted radial diffraction profile of the CsPbBr$_3$ NCs (b-e) Relative difference maps for various fluences.}
\label{fig:s3}
\end{figure}

\newpage

\begin{figure}[th!]
\renewcommand{\figurename}{Supplementary Figure}
\centering
\includegraphics[width=\linewidth]{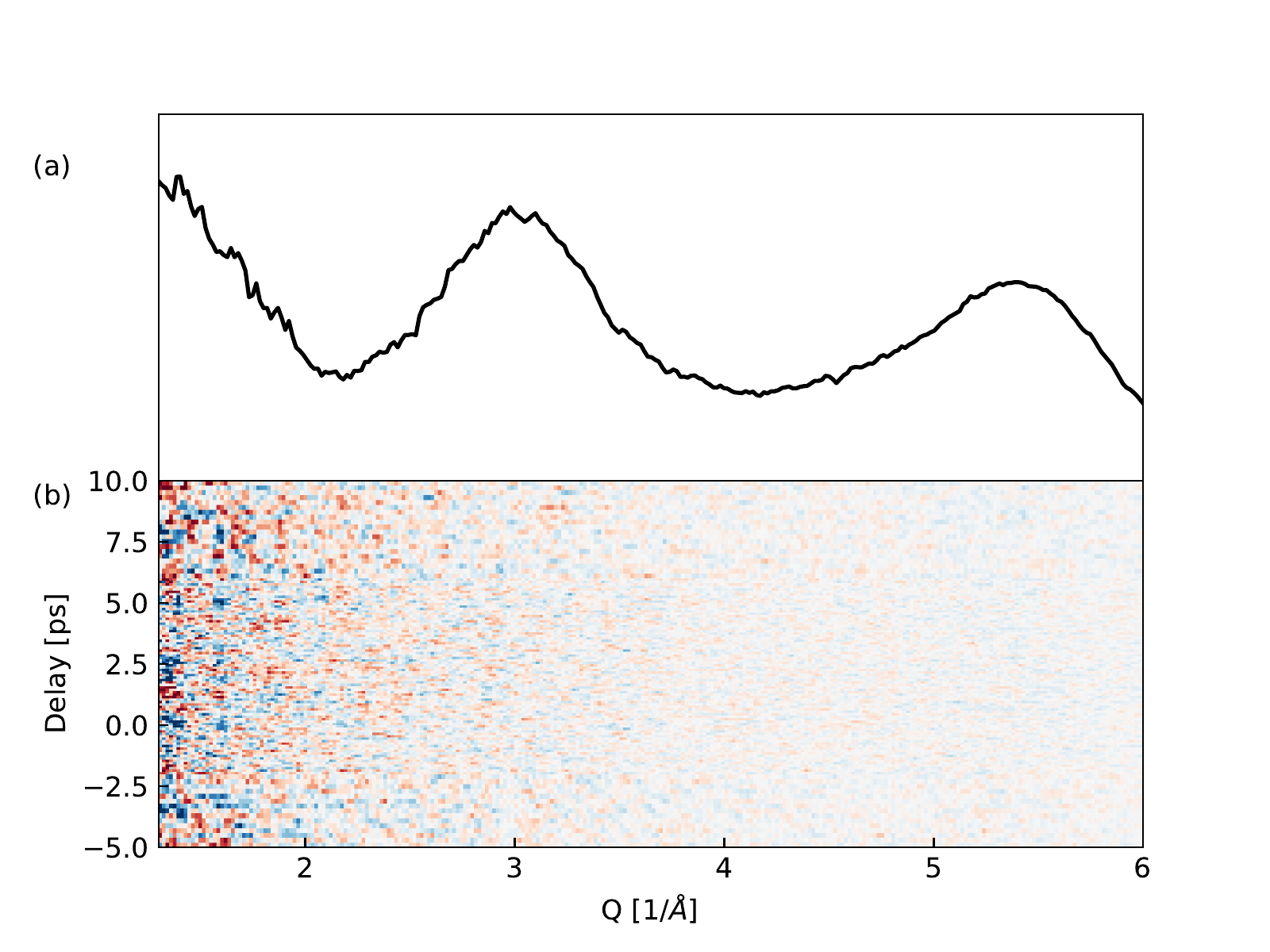}
\caption{Bare quantifoil (QF) response to photo-excitation at 3.1 eV with an incident fluence of 0.35 mJ/cm$^2$. (a) Azimuthally averaged diffraction signal of the QF membrane. Broad features can be seen around 3.0 and 5.4 [1/$\si{\angstrom}$]. (b) Difference map showing no detectable photo-induced response of the QF film.}
\label{fig:QF}
\end{figure}

\newpage

\begin{figure}[th!]
\renewcommand{\figurename}{Supplementary Figure}
\centering
\includegraphics[width=0.8\linewidth]{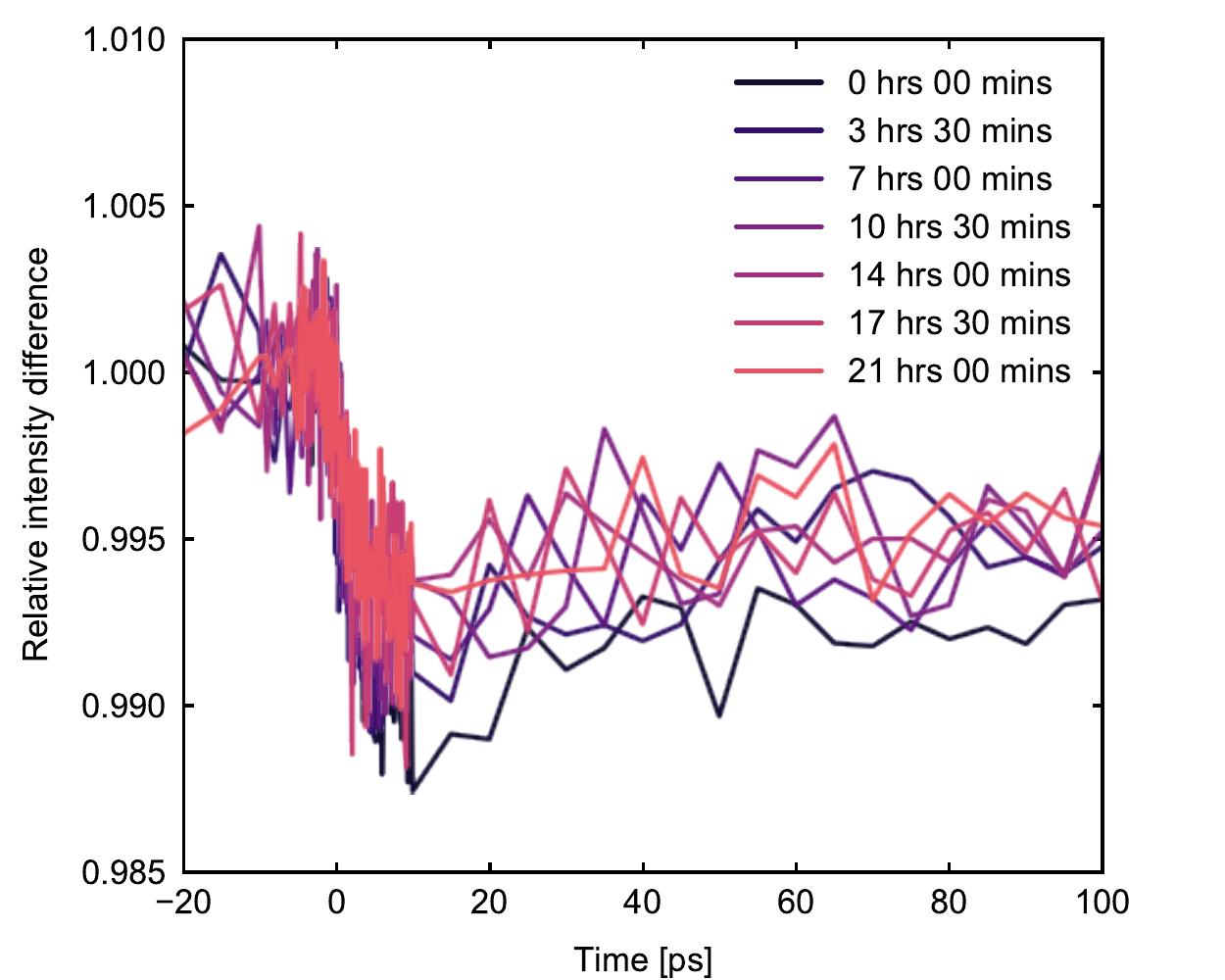}
\caption{Relative intensity changes of the diffraction signal integrated between Q = 3.50 and 3.77 $\si{\angstrom}$ (peak 5), for different repetitions of the same pump-probe delays. The caption indicates the lab time corresponding to each scan.}
\label{fig:s4}
\end{figure}

\newpage

\begin{figure}[h!]
\renewcommand{\figurename}{Supplementary Figure}
\centering
\includegraphics[width=\linewidth]{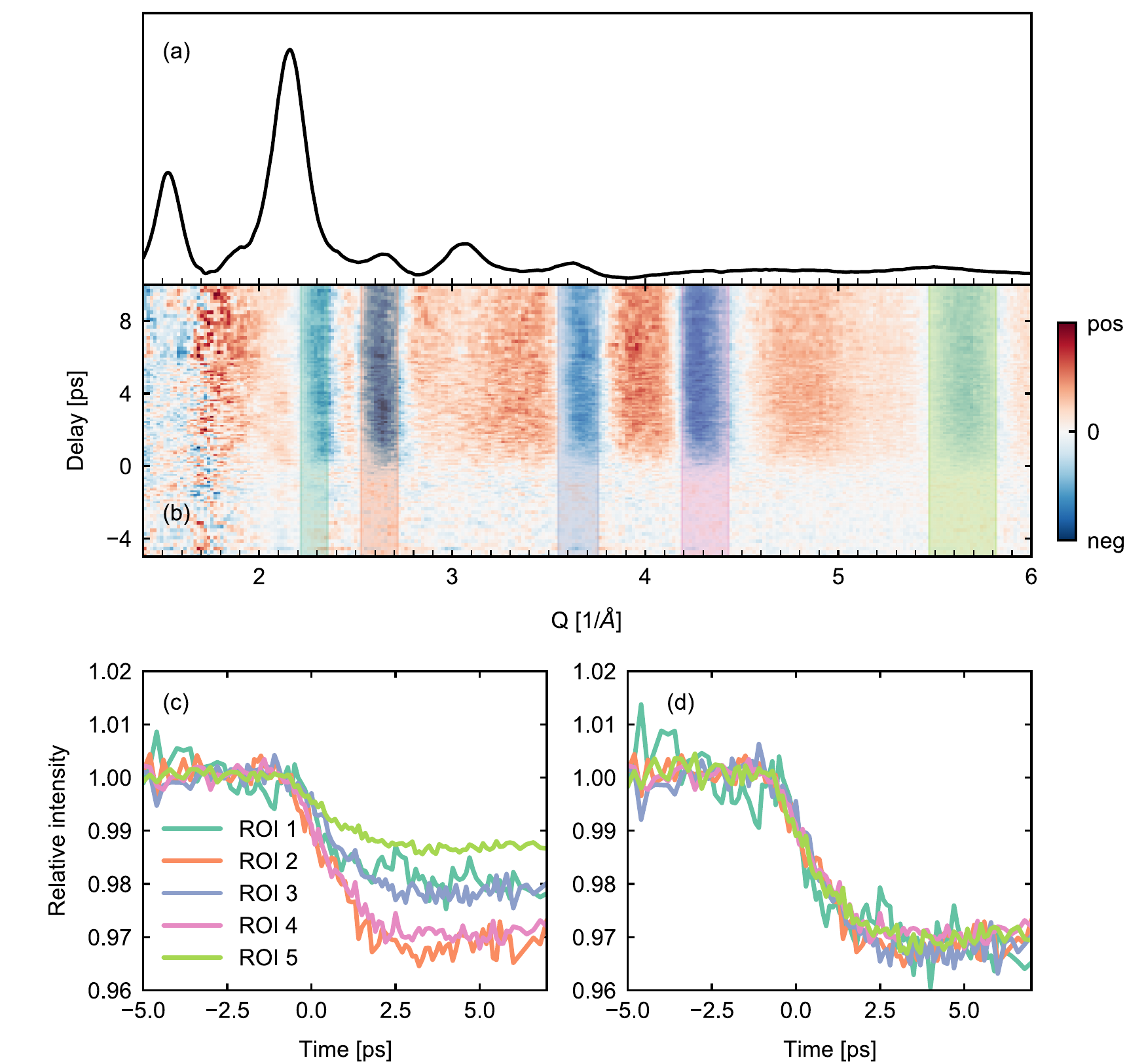}
\caption{(a-b) Each time-resolved trace shown in Figure 3 of the main text was obtained from the raw signals in the regions of interest (ROIs) marked by colored rectangles on the difference map. (c) This panel shows the average signal within each ROI, with matching colors to the rectangles in (b). (d) This panel shows the signals of panel (c) renormalized to the signal in ROI 2 (the largest signal). Since the signals between different ROIs are characterized by the same time constants within error margin, the average over the five ROIs was performed, yielding one time-resolved trace per excitation density. This trace is reported in Figure 3 of the main text.}
\label{figs6}
\end{figure}
\newpage

\begin{figure}[h!]
\renewcommand{\figurename}{Supplementary Figure}
\centering
\includegraphics[width=1\linewidth]{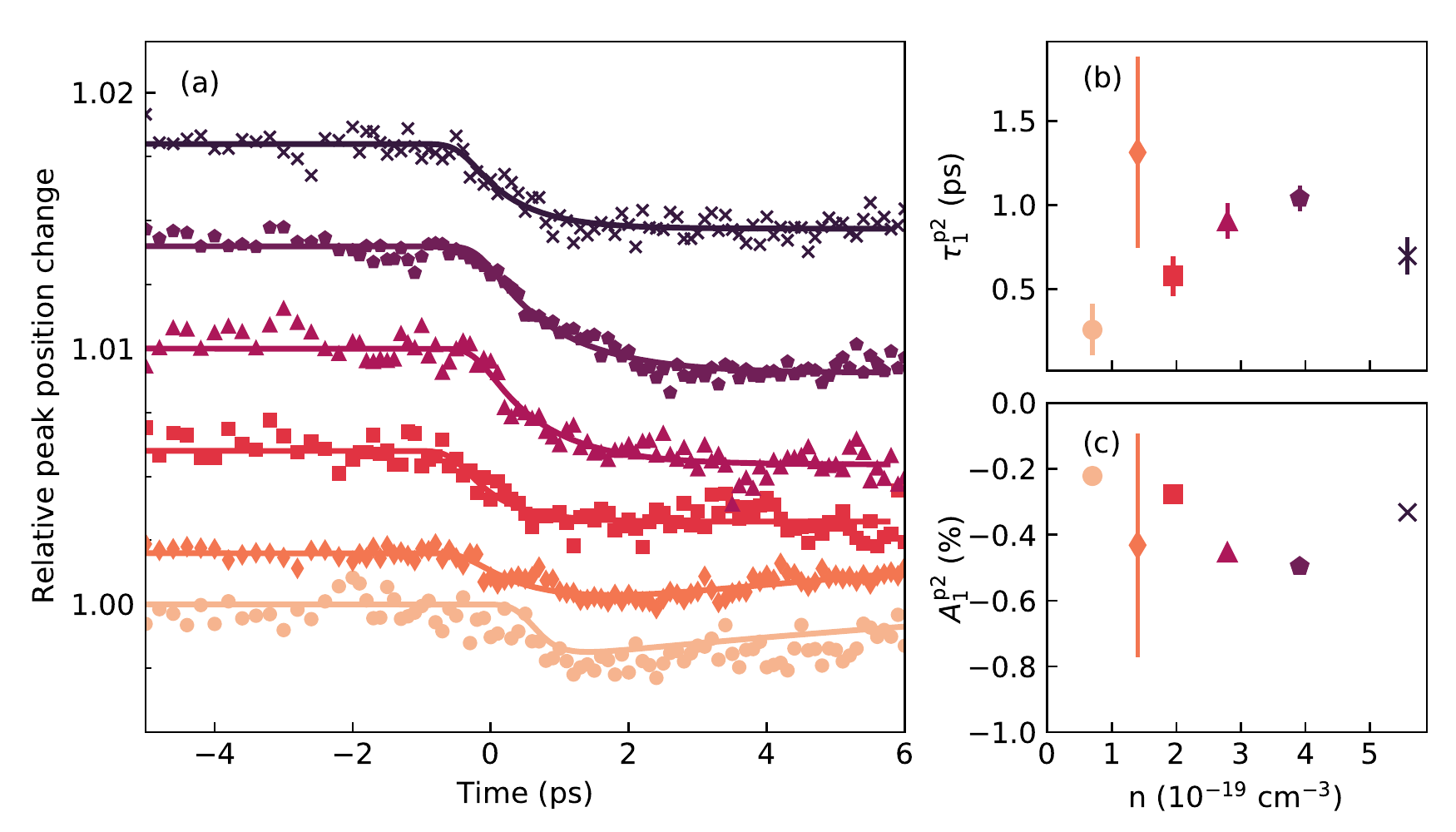}
\caption{(a) Relative peak position change of peak 2 as a function of pump-probe delay. The peak position changes were determined from changes in the peak center-of-mass, as this yielded the most reliable results. The color code is matched to that of panels (b) and (c), which display the values of the corresponding excitation densities on their x-axis. (b)  Time constant $\tau_1^{\mathrm{p2}}$ extracted from a bi-exponential fit to the data in panel (a) as a function of excitation density. (c) Amplitude $A_1^{\mathrm{p2}}$ extracted from the same fit as a function of excitation density.}
\label{figs7}
\end{figure}

\newpage

\begin{figure}[h!]
\renewcommand{\figurename}{Supplementary Figure}
\centering
\includegraphics[width=0.7\linewidth]{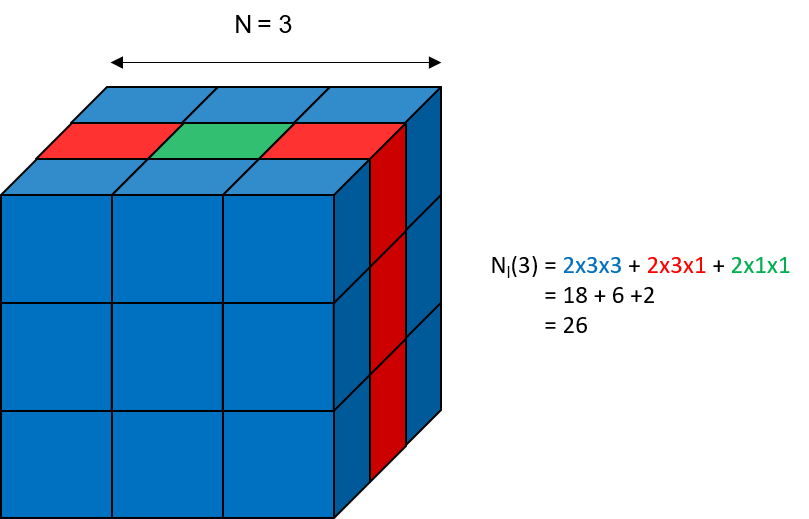}
\caption{Graphic illustration of how to determine how many unit cells, each represented by a cube, are contained in a given layer of the NC. Here illustrated for the layer surrounding the center unit cell of the NC, with an edge length of $N$ = 3 unit cells. The number of unit cells in this layer, $N_l$(3), is here 26.}
\label{fig:s3}
\end{figure}

\newpage

\begin{figure}[h!]
\renewcommand{\figurename}{Supplementary Figure}
\centering
\includegraphics[width=0.9\linewidth]{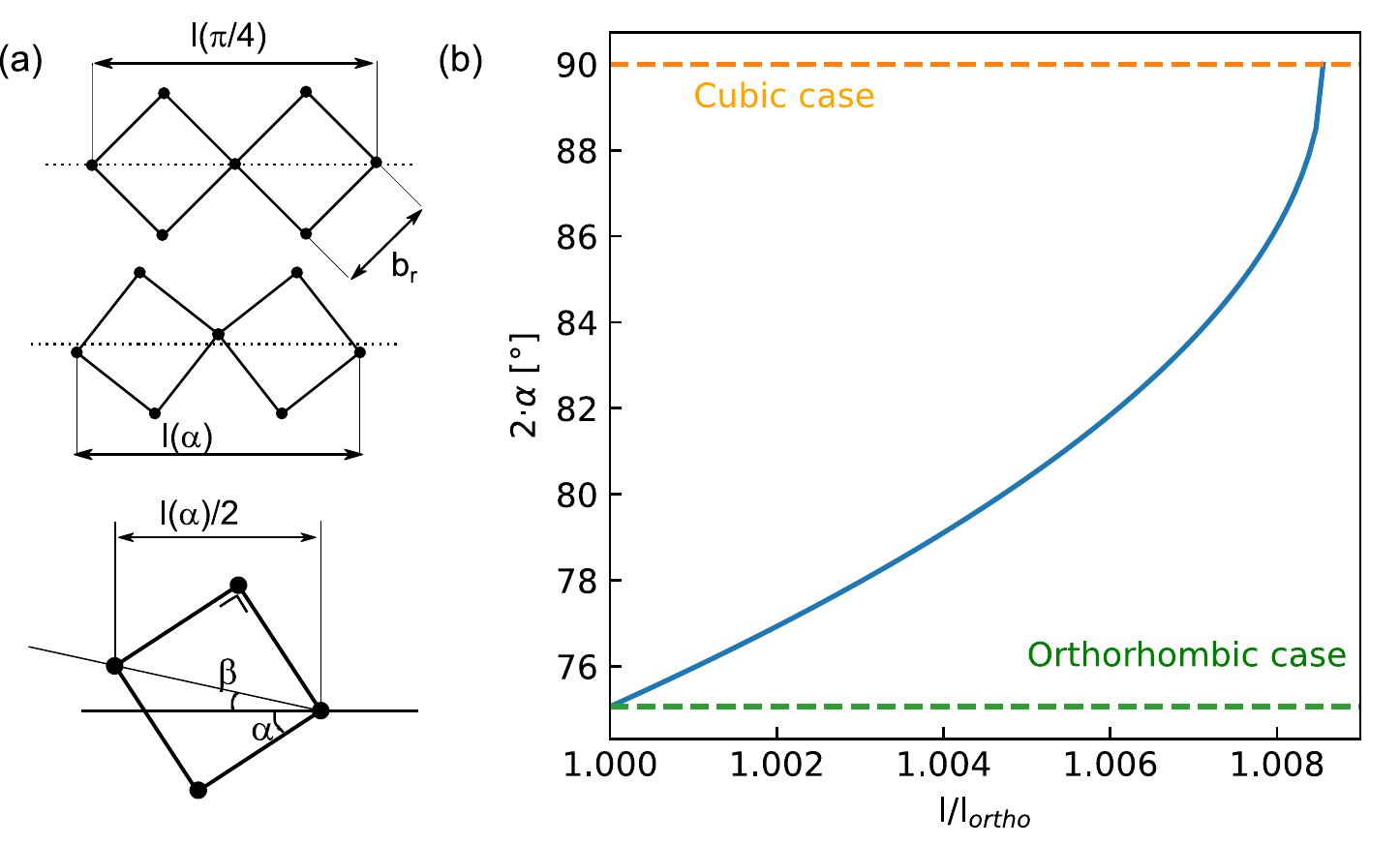}
\caption{(a) Octahedra from the perovskite lattice seen along the $b$-axis of the crystal. Only Br atoms are shown, as the black circles. The top scheme represents the cubic situation, while the middle scheme represents a case between cubic and orthorhombic. The bottom scheme defines the angles of interest. (b) Tilt angle as a function of $l/l_{\textrm{ortho}}$. We used this graph to determine the tilt angle corresponding to the shift of peak 2, which was translated in an increase of $l$.}
\label{fig:s3}
\end{figure}

\newpage

\begin{figure}[h!]
\renewcommand{\figurename}{Supplementary Figure}
\centering
\includegraphics[width=0.9\linewidth]{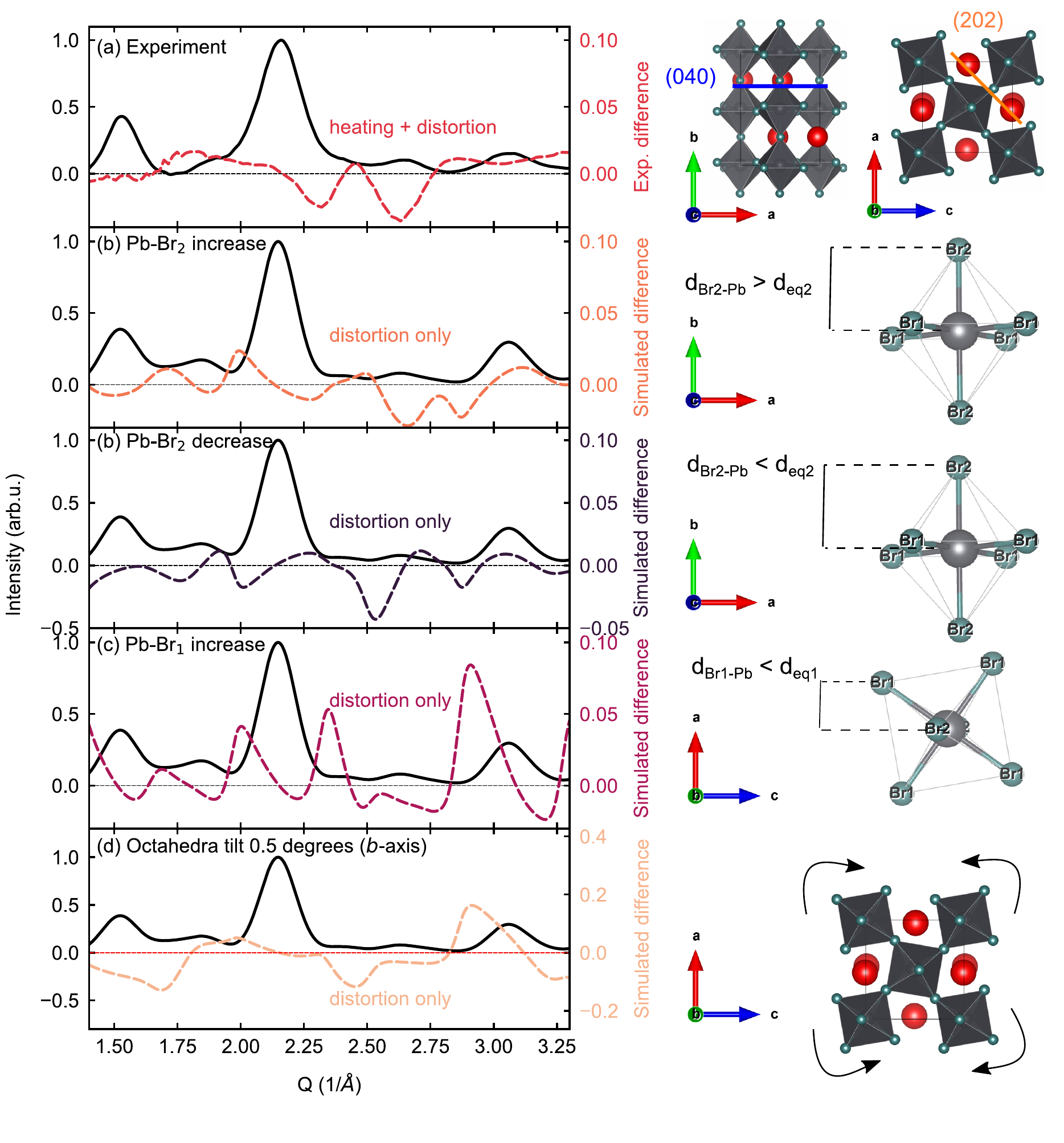}
\caption{(a) Experimental radial average (black) and relative intensity difference (dashed red) profiles. The difference signal is integrated over the late delays. The (040) and (202) Miller planes, to which peak 2 is sensitive, are shown on the right. (b-e) Exemplary simulated relative difference profiles (dashed lines) for various distortions. The corresponding distortion is shown on the right of each panel. Heating contributions, which would yield a negative contribution in the peak regions, are not included.}
\label{figs8}
\end{figure}
\newpage

\begin{figure}[h!]
\renewcommand{\figurename}{Supplementary Figure}
\centering
\includegraphics[width=0.7\linewidth]{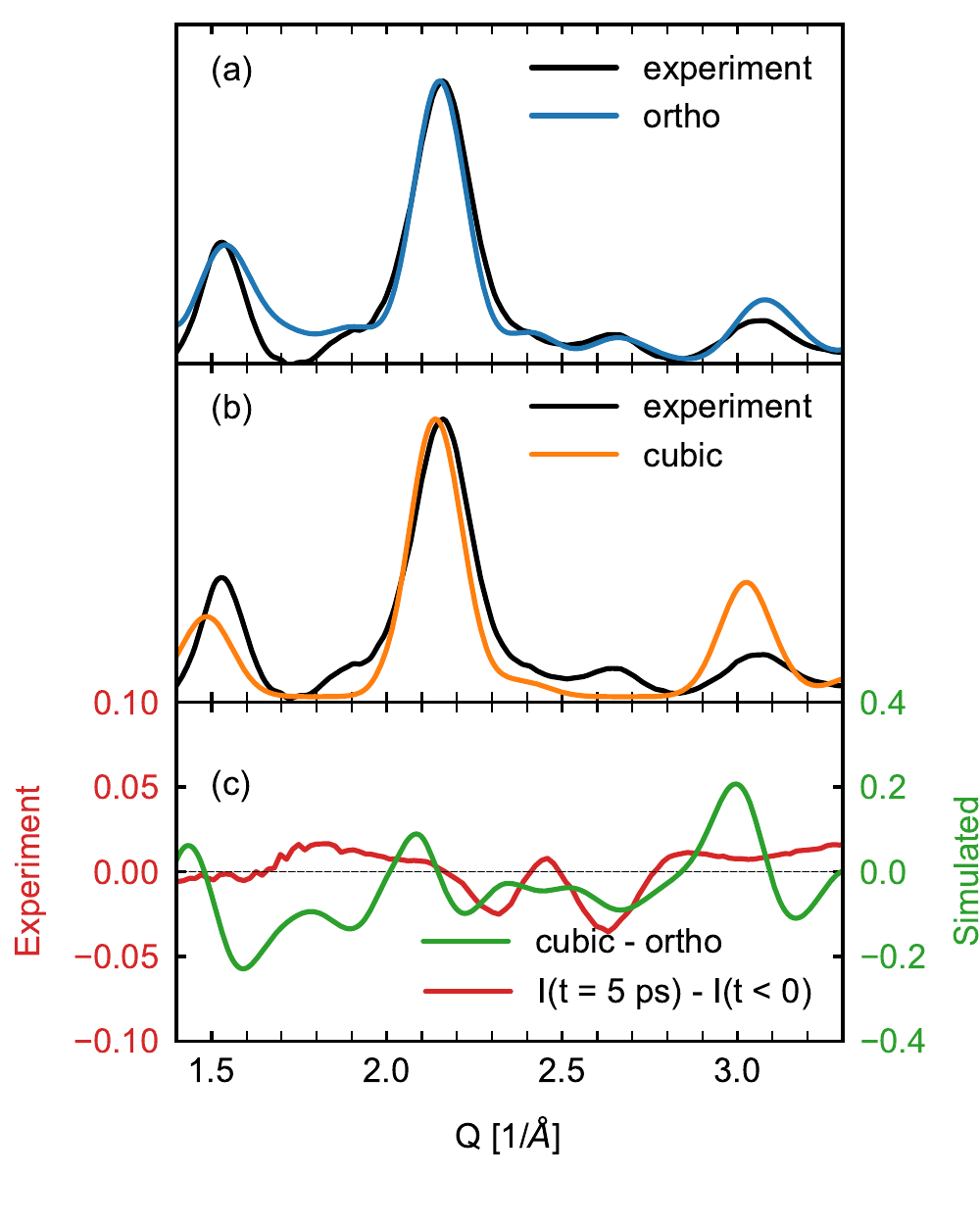}
\caption{Simulation of an orthorhombic to cubic phase transition in the CsPbBr$_3$ NCs. (a) Overlay of the simulated orthorhombic pattern and the experimental pattern. (b) Overlay of the simulated cubic pattern and the experimental pattern. (c) Pump-induced structural dynamics from the experiment (red) and difference between the cubic and orthorhombic patterns, simulating the fingerprints of a potential photo-induced phase transition.}
\label{figs7}
\end{figure}

\newpage
\section{Supplementary Tables}
\begin{table}[h!]
\centering
\begin{tabular}{l c c c c c }
Indices  & Q [1/$\si{\angstrom}$] & I/I$_{\rm max}$ [$\%$]  \\ 
\textbf{($\mathbf{1\overline{21}}$)} & \textbf{1.52} & \textbf{23.1}\\
\textbf{($\mathbf{\overline{1}2\overline{1}}$)} & \textbf{1.52} & \textbf{22.2}\\
\textbf{($\mathbf{12\overline{1}}$)} & \textbf{1.52} & \textbf{23.1}\\
\textbf{($\mathbf{121}$)} & \textbf{1.52} & \textbf{22.1}\\
\textbf{($\mathbf{200}$)} & \textbf{1.52} & \textbf{26.6}\\
\textbf{($\mathbf{002}$)} & \textbf{1.53} & \textbf{26.5}\\
($102$) & 1.71 & 4.3\\
($\overline{1}02$) & 1.71 & 3.7\\
($20\overline{1}$) & 1.71 & 3.5\\
($201$) & 1.71 & 4.0\\
\end{tabular}
\caption{Miller indices corresponding to peak 1 (1.35 < Q < 1.75 [1/$\si{\angstrom}$]), as labeled in Figure 2(a) of the main text \cite{Stoumpos2013}. Peaks are shown with increasing value of Q. Reflections below 2 $\%$ of I$_{\rm max}$ are not reported. The reflections highlighted in bold constitute the vast majority of the signal measured in peak 1.}
\label{table1}
\end{table}

\begin{table}
\begin{center}
\begin{tabular}{ c c c c c c c}
Indices  & Q [1/$\si{\angstrom}$] & I/I$_{\rm max}
$ [$\%$]  \\ 
($2\overline{2}\overline{1}$)& 2.01 & 2.9\\
($221$)& 2.01 & 3.2\\
($22\overline{1}$)& 2.01 & 2.8\\
($\overline{2}2\overline{1}$)& 2.01 & 3.3\\
($12\overline{2}$)& 2.02 & 2.7\\
($\overline{1}22$)& 2.02 & 2.8\\
($1\overline{2}2$)& 2.02 & 3.1\\
($122$)& 2.02 & 3.2\\
\textbf{($\mathbf{040}$)} & \textbf{2.14} & \textbf{100.0}\\
\textbf{($\mathbf{\overline{2}02}$)}& \textbf{2.16} & \textbf{82.6}\\
\textbf{$(\mathbf{202}$)}& \textbf{2.16} & \textbf{81.5}\\
($\overline{2}\overline{3}0$)& 2.21 & 2.3\\
($\overline{2}30$)& 2.21 & 2.4\\
($2\overline{3}0$)& 2.21 & 2.4\\
($230$)& 2.21 & 2.3\\
($\overline{1}\overline{4}1$)& 2.40 & 3.9\\
($\overline{1}4\overline{1}$)& 2.40 & 4.1\\
($1\overline{4}\overline{1}$)& 2.40 & 3.9\\
($141$)& 2.40 & 4.1\\
\label{table2}
\end{tabular}
\end{center}
\caption{Miller indices corresponding to peak 2 (1.95 < Q < 2.40 [1/$\si{\angstrom}$]), as labeled in Figure 2(a) of the main text. Peaks are shown with increasing value of Q. Reflections below 2 $\%$ of I$_{\rm max}$ are not reported. The reflections highlighted in bold constitute the vast majority of the signal measured in peak 2.}
\end{table}

\begin{table}
\begin{center}
\begin{tabular}{ c c c c c c c}
Indices  & Q [1/$\si{\angstrom}$] & I/I$_{\rm max}$ [\%]  \\ 
($\overline{2}40$)& 2.63 & 16.0\\
($240$)& 2.63 & 16.6\\
($042$)& 2.63 & 15.8\\
($04\overline{2}$)& 2.63 & 16.3\\
($321$)& 2.64 & 12.4\\
($32\overline{1}$)& 2.64 & 12.9\\
($3\overline{2}1$)& 2.64 & 12.6\\
($\overline{3}21$)& 2.64 & 13.1\\
($\overline{1}\overline{2}3$)& 2.65 & 18.3\\
($1\overline{2}3$)& 2.65 & 17.5\\
($1\overline{2}\overline{3}$)& 2.65 & 18.4\\
($123$)& 2.65 & 17.4\\
\label{table3}
\end{tabular}
\end{center}
\caption{Miller indices corresponding to peak 3 (2.5 < Q < 2.8 [1/$\si{\angstrom}$]), as labeled in Figure 2(a) of the main text. Peaks are shown with increasing value of Q. Reflections below 2 $\%$ of I$_{\rm max}$ are not reported.}
\end{table}

\begin{table}
\begin{center}
\begin{tabular}{ c c c c c c c}
Indices  & Q [1/$\si{\angstrom}$] & I/I$_{\rm max}$ [\%]  \\ 
\textbf{($\mathbf{242}$)}& \textbf{3.04} & \textbf{34.7}\\
\textbf{($\mathbf{2\overline{4}\overline{2}}$)}& \textbf{3.04} & \textbf{36.7}\\
\textbf{($\mathbf{\overline{2}4\overline{2}}$)}& \textbf{3.04} & \textbf{35.3}\\
\textbf{($\mathbf{\overline{2}\overline{4}2}$)}& \textbf{3.04} & \textbf{35.6}\\
\textbf{($\mathbf{400}$)}& \textbf{3.05} & \textbf{27.9}\\
\textbf{($\mathbf{004}$)}& \textbf{3.07} & \textbf{27.9}\\
($401$)& 3.14 & 4.4\\
($\overline{4}01$)& 3.14 & 3.4\\
($10\overline{4}$)& 3.16 & 3.4\\
($104$)& 3.16 & 4.4\\
($143$)& 3.23 & 3.3\\
($1\overline{43}$)& 3.23 & 2.8\\
($\overline{1}4\overline{3}$)& 3.23 & 3.4\\
($\overline{1}\overline{4}3$)& 3.23 & 2.8\\
($341$)& 3.23 & 5.1\\
($3\overline{4}\overline{1}$)& 3.23 & 4.8\\
($\overline{3}4\overline{1}$)& 3.23 & 5.3\\
($\overline{3}\overline{4}1$)& 3.23 & 4.6\\
\end{tabular}
\end{center}
\caption{Miller indices corresponding to peak 4 (2.85 < Q < 3.30 [1/$\si{\angstrom}$]), as labeled in Figure 2(a) of the main text. Peaks are shown with increasing value of Q. Reflections below 2 $\%$ of I$_{\rm max}$ are not reported. The reflections highlighted in bold constitute the vast majority of the signal measured in peak 4.}
\end{table}

\begin{table}
\begin{center}
\begin{tabular}{ c c c c c c c}
Indices  & Q [1/$\si{\angstrom}$] & I/I$_{\rm max}$ [\%]  \\ 
($440$)& 3.73 & 14.5\\
($4\overline{4}0$)& 3.73 & 15.1\\
($0\overline{4}4$)& 3.74 & 14.5\\
($044$)& 3.74 & 14.7\\
\end{tabular}
\end{center}
\caption{Miller indices and simulated intensities corresponding to peak 5 (3.45 < Q < 3.85 [1/$\si{\angstrom}$]), as labeled in Figure 2(a) of the main text. Peaks are shown with increasing value of Q. Reflections below 2 $\%$ of I$_{\rm max}$ are not reported. The reflections highlighted in bold constitute the vast majority of the signal measured in peak 4.}
\end{table}
\newpage
\newpage

\bibliography{bibliography}